%Paper: hep-th/9508135
%From: "Andrea Pasquinucci, Niels Bohr Institute" <pasquinu@hetws1.nbi.dk>
%Date: Fri, 25 Aug 95 11:04:18 +0200
%Date (revised): Fri, 25 Aug 95 15:31:45 +0200

%%%%%%%%%%%%%%%%%%%%%%%%%%%%%%%%%%%%%%%%%%%%%%%%%%%%%%%%%%%%%%%%%%%%%%%%
%                                                                      %
% On the Normalization and Hermiticity of Amplitudes in 4D Heterotic   %
% Superstrings, A. Pasquinucci and K. Roland, NBI-HET-95-28, August    %
% 1995, hep-th/9508135 , 1+33 pages, TeX macros included               %
%                                                                      %
% If you have difficulties in TeXing this paper or you want to TeX it  %
% in another page format, uncomment 8 lines and comment 1 line below   %
%                                                                      %
%%%%%%%%%%%%%%%%%%%%%%%%%%%%%%%%%%%%%%%%%%%%%%%%%%%%%%%%%%%%%%%%%%%%%%%%
\input panda
%
%                                PAPER
%
%
\loadamsmath
\chapterfont{\bfone} \sectionfont{\scaps}
\noblackbox

\def\Teta#1#2{\Theta\left[{}^{#1}_{#2}\right]}

\def\eqmodone{\ \buildchar{=}{{\rm{\scriptscriptstyle MOD\ 1}}}{ }\ }
\def\eqmodoneq{\ \buildchar{=}{{\rm{\scriptscriptstyle MOD\ 1}}}{?}\ }
\def\eqmodtwo{\ \buildchar{=}{{\rm{\scriptscriptstyle MOD\ 2}}}{ }\ }
\def\eqope{\ \buildchar{=}{\rm\scriptscriptstyle OPE}{ }\ }
\def\modone#1{[\hskip-1.2pt[#1]\hskip-1.2pt]}
\def\wew#1{\langle\langle\, #1\, \rangle\rangle}
\def\di{{\rm d}}
\def\bpzarrow{\ \buildchar{\longrightarrow}{{\rm{\scriptscriptstyle
              BPZ}}}{ }\ }
\nopagenumbers
{\baselineskip=12pt
\line{\hfill NBI-HE-95-28}
\line{\hfill hep-th/9508135}
\line{\hfill August, 1995}}
{\baselineskip=14pt
\vfill
\centerline{\capsone On the Normalization and Hermiticity of Amplitudes}
\sjump
\centerline{\capsone in 4D Heterotic Superstrings}
\bjump\bjump
\centerline{\scaps Andrea Pasquinucci~\footnote{$^\dagger$}{Supported
by EU grant no. ERBCHBGCT930407.} and Kaj
Roland~\footnote{$^\ddagger$}{Supported by the Carlsberg Foundation.}}
\sjump
\centerline{\sl The Niels Bohr Institute, University of Copenhagen,}
\centerline{\sl Blegdamsvej 17, DK-2100, Copenhagen, Denmark}
\bjump \vfill
%\ifdoublepage \eject\null\vfill\fi
\centerline{\capsone ABSTRACT}
\sjump
\noindent
We consider how to normalize the scattering amplitudes of 4D
heterotic superstrings in a Minkowski background. We
fix the normalization of the vacuum amplitude (the string partition
function) at each genus, and of every vertex operator describing a physical
external string state in a way consistent with unitarity of the $S$-matrix.
We also provide an explicit expression for the map relating the vertex
operator of an incoming physical state to the vertex
operator describing the same physical state, but outgoing.
This map is related to hermitean conjugation and to the
hermiticity properties of the scattering amplitudes.

\sjump \vfill
\pageno=0 \eject }
\yespagenumbers\pageno=1 \null\bjump \introsumm
String theory~[\Ref{GSW}]
remains the most promising candidate for a quantum theory
of gravity. It has also proven itself useful as a tool for perturbative
calculations in Yang-Mills theory~[\Ref{BK}].
Accordingly, it is of interest to be
able to make detailed computations of string scattering amplitudes at
any loop order. It is well known how to do this by means of
the Polyakov path integral or, equivalently, by computing vacuum
expectation values: There exists a ``master formula'' expressing
the connected part of the scattering
amplitude at each loop level as an integral
over moduli space, where the integrand is obtained as a correlation
function of vertex operators, with the appropriate
insertions of world-sheet ghosts and Picture Changing Operators
(PCOs)~[\Ref{Phong}].
In order to obtain from this ``master formula'' an actual scattering
amplitude (i.e. a number) one would have to perform the integral over
the moduli as well
as the summation over spin structures, both of which are usually
impossible by analytical means.

In this paper we address two other important points that have to be
understood in detail to be able to obtain explicit expressions for
string scattering amplitudes.

First of all, we need to know what is the correct normalization of the
vacuum amplitude (the string partition function) at each and every
genus, and also what is the normalization of all the vertex operators
describing physical external string states.
This is obviously an important issue since, for example, it is through
the proper normalization of the vertex operators that there appears the
relation between the string length scale parameter $\alpha^\prime$, the
gravitational coupling constant $\kappa$ and the gauge coupling
constants. In a second quantized theory, like Quantum Field
Theory, the proper normalizations are obtained automatically when
computing the amplitudes using, for example, Dyson's formula. Instead,
in the first quantized framework of string theory, one has to carefully
fix all normalizations in a way consistent with unitarity of the
$S$-matrix.

Second, we need to understand what is the {\sl exact} relation between the
vertex operators that we use to describe ingoing and outgoing string states
in the ``master formula''. We may formulate this more precisely:
Consider some scattering process,
$$ \lambda_{N_{out}+N_{in}} + \ldots + \lambda_{N_{out}+1}
\ \longrightarrow \ \lambda_{N_{out}} + \ldots + \lambda_{1} \ ,
\efr
where each label $\lambda$ represents a set of
single-string state quantum numbers, such as momentum, helicity, charges etc.
By definition the quantum mechanical scattering
amplitude $A_{f \leftarrow i}$ for this process is given by the $S$ matrix
element
$$ A_{f \leftarrow i} \ = \
\langle \lambda_1 , \ldots , \lambda_{N_{out}} ; in  \vert S \vert
\lambda_{N_{out}+1} , \ldots , \lambda_{N_{out}+N_{in}} ; in \rangle
\ , \nfr{transition}
that involves only ``{\it in}'' states. Incoming strings are described by
ket-states, outgoing ones by bra-states.

In string theory we may compute the connected part of this transition
amplitude by means of the ``master formula'', where each
single-string state ---whether appearing in eq.~\transition\ as a bra
or a ket--- is represented  by a vertex operator.
The question is the following: If some vertex operator
${\cal W}_{\vert \lambda \rangle}$
represents the single-string ket-state $\vert \lambda ; in \rangle$,
what is the vertex operator ${\cal W}_{\langle \lambda \vert}$
that represents the single-string bra-state
$\langle \lambda ; in \vert$?

Since $\langle\lambda; in\vert = (\vert\lambda;in\rangle )^{\dagger}$
it is clear that this question is closely related to the hermiticity
properties of the scattering amplitude: The correct choice of
${\cal W}_{\langle \lambda \vert}$ should lead to $S$-matrix elements
consistent with unitarity. In particular it should lead to tree-level
$T$-matrix elements that are real away from the momentum poles.

In field theory, in the setting of the Lehmann-Symanzik-Zimmermann
reduction formula for $S$-matrix elements,
${\cal W}_{\langle \lambda \vert}$ is just the hermitean conjugate of
${\cal W}_{\vert \lambda \rangle}$. In string theory, due primarily to
the presence of PCOs in the ``master
formula'', the relation
between ${\cal W}_{\langle \lambda \vert}$ and
${\cal W}_{\vert \lambda \rangle}$ turns out to be somewhat modified.

In practice the two problems, finding the correct normalization of the
vertex operators appearing in the ``master formula'', and deriving the exact
relation between ${\cal W}_{\langle \lambda \vert}$ and
${\cal W}_{\vert \lambda \rangle}$, can be solved at the same time.
We could imagine considering the connected part of the tree-level
two-point amplitude (i.e. the inverse propagator) and impose that this
should assume the canonical form known from field theory. But
the ``master formula'' for connected string theory amplitudes
is only well-defined on the mass shell and here the
inverse propagator vanishes identically.

Instead we consider another simple object, which is nonzero even
on-shell and just as universal as the propagator. This is the amplitude
for any given string state to emit or absorb a zero-momentum graviton
without changing any of its own quantum numbers.
More precisely we consider the term that describes the universal coupling
of gravity to the 4-momentum of
the propagating string state and the requirement that this term assumes
its canonical form yields not only an expression for the normalization
of the vertex operator in terms of the gravitational coupling $\kappa$
(which in $D=4$ dimensions is related to
Newton's constant by $\kappa^2 = 8\pi G_N$), it  also provides the
desired map between the vertex operators ${\cal W}_{\vert \lambda
\rangle}$ and ${\cal W}_{\langle \lambda \vert}$.

The procedure that we adopt is a development of the method
proposed in ref.~[\Ref{Kaj}], where it was suggested to normalize the vertex
operator of any given string state by considering the elastic scattering
of this string state, and some ``reference''
string state, for example a graviton,  at very high center-of-mass
energies, where the interactions are dominated by gravity, and require
that the tree-level amplitude for this process reproduces the
standard one dictated by the principle of equivalence. But whereas in
ref.~[\Ref{Kaj}] the method of normalization was only applied to a few
examples, in this paper we proceed to find the proper normalization for
all vertex operators in the string theory.

The relation between the vertex operators ${\cal W}_{\vert \lambda\rangle}$
and ${\cal W}_{\langle \lambda \vert}$ and the associated
question of unitarity of the $S$-matrix was also discussed in
ref.~[\Ref{Sonoda}]. We provide an explicit expression for ${\cal
W}_{\langle \lambda \vert}$,
including an overall phase factor, which
depends on the picture of the vertex operator, that was not manifest
in ref.~[\Ref{Sonoda}].

The paper is organized as follows: In section 1 we review the situation
in quantum field theory, where the Lehmann-Symanzik-Zimmermann reduction
formula involves different operators for
incoming and outgoing particles, in analogy with the situation we
encounter in string theory. In section 2 we present the ``master
formula'' for string amplitudes and section 3 contains a discussion of
the correct overall
normalization of the vacuum  amplitude. In Section 4 we obtain an ansatz
for the map between ${\cal W}_{\vert \lambda \rangle}$ and
${\cal W}_{\langle \lambda \vert}$, which is subsequently
verified in section 5, where the normalization of the vertex operators
is also derived. In Section 6 we check that our ansatz
is consistent with unitarity in the
sense that it leads to real tree-level amplitudes away from the momentum
poles. Section 7 contains an explicit example in the framework of
four-dimensional heterotic string theories built using free world-sheet
fermions. Finally we include two appendices containing various
conventions and a third appendix devoted to the proof of the compatibility
of the GSO projection and the map  between
${\cal W}_{\vert \lambda \rangle}$ and ${\cal W}_{\langle \lambda \vert}$.
\chapter{Field theory}
We can formulate scattering amplitudes in 4D field theory in a form close
to the one we use in string theory by means of the
Lehmann-Symanzik-Zimmermann reduction formula for $S$-matrix
elements~[\Ref{IZ}]:
$$
\eqalignno{& \langle \lambda_1, \ldots , \lambda_{N_{out}};
in \vert S \vert \lambda_{N_{out}+1}, \ldots,
\lambda_{N_{out}+N_{in}}, ;in
\rangle\ = \ {\rm disconnected \ terms}\ +  &\nameali{LSZvo}\cr
&\qquad \prod^{N_{out}+N_{in}}_{j=1} \left( {i \over \sqrt{Z_j}}
\right)
\int \prod_{j=1}^{N_{out}+N_{in}}
\left( {\di}^4{x}_j \right)   \langle 0 \vert {\rm T}
V_{\langle \lambda_1 \vert} (x_1)  \ldots
V_{\vert \lambda_{N_{out}+N_{in}} \rangle} (x_{N_{\rm
out}+N_{in}}) \vert 0 \rangle \ . \cr}
$$
Here we have a Field Theory Vertex (FTV)
$V_{\vert \lambda \rangle} (x)$ corresponding
to the $1$-particle ket-state $\vert \lambda ; in \rangle$
where the label $\lambda$ incorporates the $4$-momentum $p$ as well as
other quantum numbers, and similarly we have a FTV
$V_{\langle \lambda \vert} (x)$ corresponding to the 1-particle bra-state
$\langle \lambda ; in \vert$.

Since by definition of hermitean conjugation
$\langle \lambda; in \vert = (\vert \lambda ; in \rangle )^{\dagger}$,
it is not surprising that $V_{\langle \lambda \vert}(x)$
is just the hermitean conjugate of $V_{\vert \lambda \rangle } (x)$,
$$ V_{\langle \lambda
\vert}(x) = \left( V_{\vert \lambda \rangle } (x) \right)^{\dagger} \ .
\nfr{qftinoutmap}
For example, for a particle described by a real scalar field $\phi$, the
$1$-particle states are specified by their momentum only and the Field
Theory Vertices are
$$\eqalignno{ V_{\vert p \rangle} (x) & = \ e^{i p \cdot x} \left( -
\square_x + m^2 \right) \phi(x) & \nameali{exone} \cr
V_{\langle p \vert} (x) & = \ e^{-i p \cdot x} \left( -
\square_x + m^2 \right) \phi(x) \ , \cr } $$
where in both cases $p^0 = + \sqrt{\vec{p}^{\, 2} + m^2}$ and
$\square = \eta^{\mu \nu} \del_{\mu} \del_{\nu}$
with $\eta = {\rm diag} (-1,1,1,1)$.

Another example is provided by an electron with momentum $p$ and helicity
$\eta$, where
$$\eqalignno{ V_{\vert e^-,p,\eta \rangle} (x) & = \  -
\overline{\psi} (x) \left( \buildchar{\slashchar\del_x}{\leftarrow}{}
- m \right) u(\vec{p},\eta) e^{i p \cdot x} & \nameali{extwo} \cr
V_{\langle e^-,p,\eta \vert} (x) & = \ -\overline{u} (\vec{p},\eta)
\left( -\slashchar\del_x - m \right) \psi(x) e^{-i p \cdot x}  \ .
\cr} $$
Here $\overline{\psi} = \psi^{\dagger} (i \gamma^0) = \psi^{\dagger}
(-i \gamma_0)$ and $\{ \gamma^{\mu},\gamma^{\nu} \} = 2 \eta^{\mu \nu}$.
The spinor $u(\vec{p},\eta)$ of the incoming particle
with momentum $p$ and helicity $\eta$ satisfies
the Dirac equation $(i \slashchar{p} - m) u(\vec{p},\eta)=0$ and is
normalized according to
$$ u^{\dagger}(\vec{p},\eta) u(\vec{p},\eta') = 2 p^0 \delta_{\eta,\eta'} \ .
\nfr{ftspinornorm}
For particles of nonzero mass $m$ this normalization is equivalent to the
more standard one
$\overline{u}(\vec{p},\eta) u(\vec{p},\eta') = 2 m \delta_{\eta,\eta'}$,
but unlike the standard normalization condition it can also be used for
massless particles.
\chapter{String amplitudes}
In this paper we only consider 4D heterotic string models in a Minkowski
background.
We define the $T$-matrix element as the connected $S$-matrix element
with certain normalization factors removed
$$
\eqalignno{
& { \langle \lambda_1; \dots ; \lambda_{N_{out}}; in \vert S \vert
\lambda_{N_{out}+1}, \dots , \lambda_{N_{out} + N_{in}} ; in
\rangle_{\rm connected} \over
\prod_{i=1}^{N_{tot}} \left( \langle \lambda_i ; in \vert
\lambda_i ; in \rangle \right)^{1/2} }\  = & \nameali{Smatrix} \cr
&\qquad i (2\pi)^4 \delta^4 ( p_1+ \dots + p_{N_{out}} - p_{N_{out}+1} -
\dots - p_{N_{tot}} ) \ \prod_{i=1}^{N_{tot}}
(2 p_i^0 V)^{-1/2} \ \times \cr
&\qquad \  T ( \lambda_1; \dots ; \lambda_{N_{out}} \vert
\lambda_{N_{out}+1}; \dots ; \lambda_{N_{out} + N_{in}} ) \ , \cr }
$$
where $N_{tot} = N_{in} + N_{out}$ is the total number of
external states, $p_i$ is the momentum of the $i$'th string state, all
of them having $p_i^0 > 0$, and $V$ is the
usual volume-of-the-world factor.  We also introduce the dimensionless
momentum $k_\mu \equiv \sqrt{{\alpha^\prime \over 2}} \, p_\mu$.
The Minkowski metric is $\eta = {\rm diag}(-1,1,1,1)$.

For heterotic superstrings in the Neveu-Schwarz Ramond formalism we have
various free conformal fields: The space-time coordinates $X^{\mu}$,
their chiral world-sheet superpartners $\psi^{\mu}$, the
reparametrization ghosts $b,c$ and $\bar{b},\bar{c}$, and the
superghosts $\beta,\gamma$. On top of this we have various internal
degrees of freedom described by a conformal field theory (CFT) with
left-moving (right-moving) central charge 22 (9). These may or may not
be free.
The $g$-loop contribution to the
$T$-matrix element is given by the Polyakov path
integral which is equivalent to the
following operator formula
$$
\eqalignno{ & T^g
(\lambda_1; \dots ; \lambda_{N_{out}} \vert
\lambda_{N_{out}+1};\dots;\lambda_{N_{out} + N_{in}} ) \ =
& \nameali{Tmatrix} \cr
&\qquad  (-1)^{g-1} C_g \int \prod_{I=1}^{3g-3+N_{tot}}
\left( \di^2 m^I \right) \ \prod_{\mu=1}^{g}
\left(\sum_{{\bfmath\alpha}_{\mu},{\bfmath\beta}_{\mu}}
C^{{\bfmath\alpha}_{\mu}}_{{\bfmath\beta}_{\mu}} \right) \
\wew{\left| \prod_{I=1}^{3g-3+N_{tot}} (\eta_I \vert b)
\prod_{i=1}^{N_{tot}} c(z_i) \right|^2\ \times \cr
&\qquad \left(\prod_{A=1}^{2g-2+N_B+N_{FP}}
\Pi (w_A)\right) \, {\cal V}_{\langle \lambda_1 \vert } (z_1,\bar{z}_1)
\dots {\cal V}_{\vert \lambda_{N_{tot}} \rangle}
(z_{N_{tot}},\bar{z}_{N_{tot}})}  \ . \cr}
$$
Here $C_g$ is a constant giving the proper normalization to the string
partition function (the $g$-loop vacuum amplitude). It will be given
explicitly in section 3, and (as we shall see) the sign
$(-1)^{g-1}$ ensures that $C_g$ is a positive number.
$m^I$ is a modular parameter,
$\eta_I$ is the corresponding Beltrami
differential, and our conventions for the overlap $(\eta_I \vert b)$
with the antighost
field $b$ are defined in detail in ref.~[\Ref{Kaj2}].
The integral is over one
fundamental domain of $N_{tot}$-punctured genus $g$ moduli space.
For each loop, labelled by $l=1,\ldots,g$, we have a summation over
sets of spin structures, collected in vectors ${\bfmath\alpha}_l$ and
${\bfmath\beta}_l$, and with a summation coefficient
$C^{{\bfmath\alpha}_l}_{{\bfmath\beta}_l}$.
By definition the correlator $\wew{\dots}$ includes the partition
function. At tree level, where the non-zero mode
partition function is equal to one,
the notation $\langle \dots \rangle$ is also used. At loop level we
choose the normalization for the partition function to be the one
obtained by applying the sewing procedure. This guarantees sensible
factorization properties in the corner of moduli space
where the world-sheet degenerates into individual tori connected by long
tubes and implies that the spin-structure summation coefficient is just
a product of one-loop summation coefficients as in eq.~\Tmatrix .
More details on our conventions for spin structures, partition functions and
operator fields in the explicit setting of a heterotic string model
built with free world-sheet fermions~[\Ref{KLT},\Ref{Anto},\Ref{Bluhm}]
can be found in Appendix A, see also refs.~[\Ref{ammedm},\Ref{mink}].

In analogy with field theory we have introduced a vertex operator
${\cal V}_{\vert \lambda \rangle} (z,\bar{z})$ for each ket string state
$\vert \lambda \rangle$ and similarly a vertex operator ${\cal
V}_{\langle \lambda \vert} (z,\bar{z})$ corresponding to each bra string
state $\langle \lambda \vert$.~\note{Since all the states we consider
are of the ``{\it in}'' variety, we drop the ``{\it in}'' label from now
on.} At the end of this section we will have more to say about the
meaning of these operators.

The ghost factors residing in the BRST invariant version of the
vertex operator, given by
$$ {\cal W}_{\vert \lambda \rangle} (z,\bar{z})\ =\
c(z) \overline{c}(\bar{z})
{\cal V}_{\vert \lambda \rangle} (z,\bar{z}) \qquad {\rm and} \qquad
{\cal W}_{\langle \lambda \vert} (z,\bar{z})\ =\
c(z) \overline{c}(\bar{z})
{\cal V}_{\langle \lambda \vert} (z,\bar{z}) \ ,
\nfr{wvvect}
have been factored out in eq.~\Tmatrix. We take all
space-time bosonic vertex operators to be in the
$q=-1$ superghost picture and all the space-time fermionic vertex
operators to be in the $q=-1/2$ superghost picture.
In an amplitude
involving $N_B$ space-time bosons and $2N_{FP}$
space-time fermions this implies that we have to insert $2g-2+N_B + N_{FP}$
PCOs $\Pi$ at
arbitrary points $w_A$ on the Riemann surface. In practical calculations
it can be convenient to insert one PCO at each of the vertex operators
describing the space-time bosons so as to change these into the $q=0$
picture. This leaves $2g-2+N_{FP}$ PCOs at
arbitrary points.

If we ``bosonize'' the superghosts in the usual way,
$\beta = \partial \xi e^{-\phi}$ and $\gamma = e^{+\phi}
\eta$, the PCO is given explicitly by
$$ \Pi = 2 c \partial \xi + 2 e^{\phi} T_F^{[X,\psi]} - {1 \over 2}
\partial (e^{2\phi} \eta b) - {1 \over 2} e^{2\phi} (\partial \eta) b \ ,
\nfr{PCO}
where we suppressed the superghost cocycle factor which ensures that
$e^{\phi}$ anti-commutes with all other fermionic operators on the
world-sheet, and
$$ T_F^{[X,\psi]} \ = \ - {i \over 2} \del X \cdot \psi + ({\rm internal
\ part }) \nfr{supcurr}
is the orbital part of the world-sheet
supercurrent (i.e. the part not involving ghosts and superghosts).
The ``internal part'' refers to the internal right-moving degrees
of freedom of the CFT with central charge $9$.

As stated in the introduction our aim in this paper is twofold:
First, since the $T$-matrix element as defined in eq.~\Smatrix\
corresponds to the connected $S$-matrix element obtained using states with
standard field theory normalization, we have to use vertex operators
with a definite normalization in eq.~\Tmatrix . So we need to know what
is the correct normalization of all vertex operators involved in the
theory; and we also need to determine the value of the overall
normalization constant $C_g$.
Second, we need to understand what is the {\sl exact} relation between the
vertex operators
${\cal W}_{\langle \lambda \vert}(z,\bar{z})$ and
${\cal W}_{\vert \lambda \rangle}(z,\bar{z})$.
By definition the operator ${\cal W}_{\vert
\lambda \rangle} (z=0)$, when acting on the conformal vacuum $\vert 0
\rangle$, creates the string state $\vert \lambda \rangle$, where (like
in section 1) $\lambda$ is a label incorporating the $4$-momentum $k$ (with
$k^0 > 0$), the helicity and the ``particle type'' (defined through the
values of various charges and family labels).
We may think of eq.~\Tmatrix\ as an indirect definition of what we mean
by ${\cal W}_{\langle \lambda \vert}$: It is the vertex operator we have
to use on the right-hand side of this equation in order to obtain the
$T$-matrix element involving the bra-state $\langle \lambda \vert =
(\vert \lambda \rangle)^{\dagger}$. Of course this definition is
somewhat circular, because we don't know how to compute the $T$-matrix
element until we have specified what are the vertex operators.
Indeed, as explained in the introduction,
the procedure we adopt is to carefully {\it derive}
what ${\cal W}_{\langle \lambda \vert}$
should be in order for the ``master formula'' \Tmatrix\ to reproduce
the correct amplitude for a propagating string to emit or
absorb a zero-momentum graviton.
Based on our experience from field theory, as outlined in section 1, we
might expect ${\cal W}_{\langle \lambda \vert}$ to be given by the
hermitean conjugate of ${\cal W}_{\vert \lambda \rangle}$. As we shall
see in section 4, this is not completely correct.
\chapter{Normalization of the vacuum amplitude}
In section 2 we already made use of the basic fact that
the problem of normalizing string amplitudes
can be separated into two independent problems: One, to fix the
normalization constant $C_g$ of the vacuum amplitude at genus $g$. The
other, to fix the normalization of each vertex operator in the theory.

It is factorization that leads to this simple result. For example, to
see that the normalization of the vertex operators cannot depend on
the topology of the world-sheet we can imagine
inserting a vertex operator on a sphere connected by a long tube to some
genus $g$ surface. It is clear that the vertex operator cannot
know about the distant handles. This is true even for vertex operators
describing space-time fermions, even though these involve spin fields
which are non-local operators on the world-sheet, because space-time
fermions always come in pairs and we may imagine isolating both of the
corresponding vertex operators (and the branch cut connecting them) on
a sphere far away from all handles.

Similarly, if we assume for the moment that the overall normalization of
the amplitude depends on the number $N$ of external states,~\note{In this
section only we drop the label {\it tot} on $N_{tot}$.}
as well as on the genus $g$, through some
coefficients $C_{g,N}$, we find by factorizing
the $N$-point $g$--loop amplitude into an $N+1$-point $g_1$--loop amplitude
times a $1$-point $g_2$--loop
amplitude times a propagator (where $g_1+g_2=g$), that
$$
C_{g_1+g_2,N}\ \propto\ C_{g_1,N+1}\ C_{g_2,1}
\efr
with a proportionality constant independent of $g_1$, $g_2$ and
$N$. Setting $g_1=0$ one gets
$$
C_{g,N}\ \propto\ C_{0,N+1}\ C_{g,1} \ ,
\efr
so that the dependence on $N$ can be studied at tree level.
Again by factorization, at tree level one gets
$$
C_{0,N_1+N_2}\ \propto\ C_{0,N_1+1}\ C_{0,N_2+1} \ ,
\efr
and if we put $N_2=2$ this implies that the ratio $C_{0,N+2} / C_{0,N+1}$
is independent of $N$
or, in other words, that $C_{0,N} \propto ({\cal M})^{N}$ for some constant
${\cal M}$. So we may write
$$
C_{g,N}\ =\ C_g\, ({\cal M})^{N} \ ,
\efr
and if we absorb a factor of ${\cal M}$
into the normalization of all vertex operators we are then left with an
overall normalization constant $C_g$ depending only on the genus.

To determine the value of $C_g$ we adopt the method proposed in
refs.~[\Ref{Kaj},\Ref{Marco}]:
To consider the elastic scattering of two gravitons in
the Regge regime of very high center-of-mass energy and small energy
transfer and impose that
the leading part of the $g$-loop amplitude assumes the universal form
needed for the eikonal resummation [\Ref{Veneziano}].

In order to get started we need the expression for the graviton vertex
operator including the proper normalization which was found in
refs.~[\Ref{Weinberg},\Ref{Kaj}]:
$${\cal V}^{(-1)}_{\vert {\rm grav} \rangle} (z,\bar{z}) = i
{\kappa \over \pi}
\bar{\epsilon} \cdot \bar{\del} X (\bar{z}) \epsilon \cdot \psi (z)
e^{-\phi(z)} e^{i k \cdot X(z,\bar{z})} \ , \nfr{gravvert}
where $k^2=0$ and we wrote the graviton polarization on the factorized
form $\bar{\epsilon} \otimes \epsilon$ with $\epsilon \cdot k =
\bar{\epsilon} \cdot k = 0$.
Our conventions for the operator fields can be found in Appendix A.
Like in eq.~\PCO\ we suppressed the cocycle factor which ensures that
the superghost
operator $e^{-\phi} = \delta(\gamma)$ anticommutes with all other
fermions on the world-sheet.

By picture changing \gravvert\ we arrive at
$$\eqalignno{  {\cal V}_{\vert {\rm grav} \rangle}^{(0)}
(z,\bar{z}) & = \ \lim_{w
\rightarrow z} \Pi(w) {\cal V}^{(-1)}_{\vert {\rm grav}
\rangle} (z,\bar{z}) \  &
\nameali{gravverttwo} \cr
& = \ {\kappa \over \pi} \bar{\epsilon} \cdot \bar{\del} X (\bar{z}) \left[
\epsilon \cdot \del X (z) - i k \cdot \psi (z) \epsilon \cdot \psi (z)
\right] e^{i k \cdot X (z,\bar{z})} \ . \cr } $$
The expressions for ${\cal V}^{(-1)}_{\langle {\rm grav} \vert}$ and
${\cal V}^{(0)}_{\langle {\rm grav} \vert}$ are identical to
eqs.~\gravvert\ and \gravverttwo , as long as the polarizations
$\epsilon,\bar{\epsilon}$ are taken to be real, and we ascribe to the
outgoing graviton a momentum with $k^0 < 0$.

The calculation of the four-graviton $g$-loop amplitude in the Regge
limit starting from eq.~\Tmatrix\
is different from the one in ref.~[\Ref{Kaj}] which was performed using
the manifestly world-sheet supersymmetric formulation of the heterotic
string.
In fact it is much harder, because even after changing the graviton
vertex operators into the $(0)$ picture there remains $2g-2$ PCOs at
arbitrary points. To obtain the universal form of the amplitude in the
pinching limit relevant for the Regge regime, where the world-sheet
degenerates into a ladder-like configuration consisting of two ``fast
legs'' connected by $g+1$ long tubes, one should insert $g-1$ PCOs on
each of the two ``fast legs''. (Other choices are of course possible but
will lead to the presence of total derivatives that make the leading
behaviour of the amplitude rather obscure.) Even subject to this
constraint there still remains $2g-2$ PCO insertion points, the
dependence on which only drops out at the very end of the calculation.

In the end we recover the standard result [\Ref{Kaj}] pertaining to $D=4$
space-time dimensions,
$$ C_g = \left( { 2\kappa^2 \over \alpha' } \right)^{g-1} \left( {1
\over 2\pi } \right)^{5g-3} (\alpha')^{-2} \nfr{overallnorm}
{\it and} the sign factor $(-1)^{g-1}$ explicitly displayed in
eq.~\Tmatrix. The origin of this sign is not too hard to understand. It is
needed to compensate the identical sign which
appears when we disentangle the anticommuting superghost factors
$e^{\phi}$ and
the orbital supercurrents $T_F^{[X,\psi]}$ in the product of the $2g-2$
PCOs
$$ \prod_{\alpha=1}^{2g-2} \left( e^{\phi(w_{\alpha})} T_F^{[X,\psi]}
(w_{\alpha}) \right) = (-1)^{g-1} \left(
\prod_{\alpha=1}^{2g-2} e^{\phi(w_{\alpha})} \right)
\left( \prod_{\alpha=1}^{2g-2} T_F^{[X,\psi]}
(w_{\alpha}) \right) \ . \efr
The other three terms present in the PCO \PCO\ do not contribute to the
leading behaviour of the amplitude in the Regge regime.

A comment about the spin structure summation coefficient
in eq.~\Tmatrix\ might be in order at
this point: We fix $C_g$ by considering the four-graviton $g$-loop
amplitude in the Regge regime. However, only the $2^g$ spin structures
responsible for graviton exchange contribute to the
leading, universal part of the amplitude.
How do we know that the
normalization we obtain is also correct for all the other spin
structures? The answer to this has already been given in section 2: The
requirement that the amplitude factorizes properly in the limit where
all loops are taken far apart implies that the spin structure summation
coefficient should be a product of one-loop summation coefficients.
These are in turn specified by the requirement that the one-loop
partition function should be modular invariant, once a (physically
sensible) choice of GSO projection has been
made~[\Ref{KLT},\Ref{Anto}].~\note{Strictly speaking modular invariance
of the one-loop partition function does not specify the summation
coefficient for those spin structures where one (or more) of the free
fermions on the world-sheet develop a zero mode, because these spin
structures give zero contribution to the partition function. In order to
check that no extra phase factors appear in these cases one may for
example consider the factorization of a two-loop vacuum amplitude into
one-loop tadpoles~[\Ref{Anto}]. We carried out this check explicitly in
the framework of Kawai-Lewellen-Tye~[\Ref{KLT}] heterotic string models.}
\chapter{The relation between ${\cal W}_{\vert \lambda \rangle}$ and
${\cal W}_{\langle \lambda \vert}$ }
We now consider in detail the connection between the vertex operators
describing incoming and outgoing string states.

What we are looking for is the map which, given the vertex operator
${\cal W}_{\vert \lambda \rangle}$
describing an incoming string state, gives us the vertex operator
${\cal W}_{\langle \lambda \vert}$
describing the same string state but outgoing. As we saw in section 1
this map is just given by hermitean conjugation in the framework of
quantum field theory. In string theory this cannot be the whole story,
because if the operator field ${\cal W}_{\vert \lambda \rangle}$
creates the ket-state $\vert \lambda \rangle$ in the usual sense,
$\vert \lambda \rangle =\lim_{\zeta,\bar{\zeta}\rightarrow 0}
{\cal W}_{\vert \lambda \rangle} (\zeta,\bar{\zeta}) \vert 0 \rangle $,
then by definition the
hermitean conjugate operator field creates the corresponding bra-state,
$\langle \lambda \vert =
\lim_{\zeta,\bar{\zeta}\rightarrow 0} \langle 0\vert
\left( {\cal W}_{\vert \lambda \rangle} (\zeta,\bar{\zeta})
\right)^\dagger$.
But in eq.~\Tmatrix\ both ${\cal V}_{\langle \lambda \vert}$ and
${\cal V}_{\vert \lambda \rangle}$ are vertex operators that create
ket states when acting on the conformal ket vacuum.

So we need to compose two-dimensional hermitean conjugation
with some other transformation which also maps a vertex operator
creating ket-states into a vertex operator creating bra-states.
This transformation should be a symmetry of any 2-dimensional conformal
field theory on the sphere. The obvious choice is the
{\sl BPZ conjugation\/} [\Ref{BPZ}] (see also [\Ref{Zwiebach}]).

Therefore we now quickly review our conventions on hermitean
conjugation and BPZ conjugation in conformal field theory.
After that we will propose a map from ${\cal W}_{\vert \lambda \rangle}$
to ${\cal W}_{\langle \lambda \vert}$
which is just an unknown phase
factor times the combination of
BPZ and hermitean conjugation. In the next section we will check
that our guess indeed gives the right map, and in the process the phase
factor will be determined.
\section{Two-dimensional hermitean conjugation}
In this section we review our conventions on hermitean conjugation, see
also refs.~[\Ref{Sonoda},\Ref{mink}]. We define the hermitean conjugate of all
elementary
operators in the conformal field theory by specifying the hermitean
conjugate of the corresponding oscillators, with the further
understanding that hermitean conjugation also complex conjugates all
complex numbers and inverts the order of the operators.

For example, if
$$\Phi_{\Delta} (z) = \sum_n \phi_n z^{-n-\Delta} \efr
is a primary chiral conformal field of conformal dimension $\Delta$, then the
hermitean conjugate of this field is
$$
\left(\Phi_{\Delta}(z)\right)^\dagger\ =\
\left({1\over z^*}\right)^{2\Delta}
\widehat\Phi_{\Delta}
({1\over z^*})\ ,
\nfr{hermconj}
where $z^*$ denotes the complex conjugate of $z$ (we think of $z$
and $\bar{z}$ as independent complex variables, so that $z^*$ and $\bar{z}$
need not be equal) and
$$ \widehat\Phi_{\Delta} (z) = \sum_n \phi_{-n}^{\dagger} z^{-n-\Delta}
\nfr{hermmode}
is a primary conformal field of the same dimension as $\Phi_{\Delta}$.
We say that a field $\Phi_{\Delta}$ is hermitean (anti-hermitean)
when $\widehat\Phi_{\Delta} = +\Phi_{\Delta} \ (-\Phi_{\Delta})$.

The hermiticity properties are made more complicated by the presence of
the reparametrization ghosts, because on the sphere the basic
nonvanishing correlator is
$ \langle \bar{c}_{-1} \bar{c}_0 \bar{c}_{1} c_{-1} c_0 c_1 \rangle $
where (since $c_n^{\dagger} = c_{-n}$)
the operator involved is explicitly anti-hermitean. Therefore
either one has to postulate an imaginary value for this correlator or
one has to relinquish the property $\langle M \vert A \vert N \rangle =
+ \langle N \vert A^{\dagger} \vert M \rangle^*$ of matrix elements
involving ghost degrees of freedom. We prefer the second option. We define
$$\langle \ \vert c_{-1} c_0 c_1 \vert^2 \ \rangle = \langle
\bar{c}_{-1} \bar{c}_0 \bar{c}_{1} c_{-1} c_0 c_1 \rangle = +1
\nfr{ghostone}
and this implies that
$$ \langle M \vert A \vert N \rangle = - \langle N \vert A^{\dagger}
\vert M \rangle^* \nfr{ghostherm}
in the presence of ghosts. As a special case of this
$$ \langle M \vert c_0 \bar{c}_0 A \vert N \rangle = \langle N \vert c_0
\bar{c}_0 A^{\dagger} \vert M \rangle^* \nfr{ghosttwo}
for any operator $A$ not involving the modes $b_0$ or $\bar{b}_0$.

A list of hermiticity properties for the fields relevant in
four-dimensional heterotic string models constructed using free fermions
can be found in Appendix B.
\section{BPZ invariance in conformal field theories}
Consider a conformal field theory on the cylinder.
Introduce complex coordinates $z=\exp\{ i (\sigma+\tau)\}$ and
$\bar{z} = \exp\{i(-\sigma+\tau)\}$ and rotate
to Euclidean time $\tau \rightarrow -i \tau$.
Changing sign on $\tau$ and $\sigma$ simultaneously gives rise to the
Belavin-Polyakov-Zamolodchikov (BPZ) transformation $z \rightarrow 1/z$
[\Ref{BPZ},\Ref{Zwiebach}]. This transformation
defines a globally holomorphic diffeomorfism on the sphere.

At the level of the operator
fields, the transformation changes the coordinate system from $(z)$ to
$(w)$ where $w=1/z$:
$$ \Phi (z=\zeta) \bpzarrow \Phi (w=\zeta) \ .
\nfr{bpz}
For a primary conformal field of dimension $\Delta$
$$ \Phi_{\Delta} (w=\zeta) =
e^{-i \epsilon \pi \Delta} \left( {1 \over \zeta} \right)^{2\Delta}
\Phi_{\Delta} \left( z= {1 \over \zeta} \right) \ , \nfr{bpzprimary}
where for non-integer conformal dimensions we have to
choose a specific phase for $-1$,~\note{In
their original paper~[\Ref{BPZ}], Belavin, Polyakov and
Zamolodchikov avoided this problem by considering instead the conformal
transformation $z \rightarrow -1/z$, but we prefer to consider
$z \rightarrow 1/z$, in accordance with most subsequent authors.}
parametrized by an odd integer
$\epsilon$, when forming the transformation factors
$$ {dz \over dw} = e^{-i \epsilon \pi} {1 \over w^2} \qquad {\rm and}
\qquad
{dw \over dz} = e^{+i \epsilon \pi} {1 \over z^2} \ . \nfr{transfac}
The BPZ transformation does not reverse the order of operators and it
leaves all complex numbers unchanged. It cannot itself be generated by
any operator acting on ket states.
Instead it defines a map from ket-states to bra-states as follows:
$$ \vert \Phi \rangle \ \equiv \ \lim_{\zeta \rightarrow 0}
\Phi (z=\zeta) \vert 0 \rangle \ \ \bpzarrow \ \
\langle \Phi^{\rm BPZ} \vert \ \equiv \ \lim_{\zeta \rightarrow 0}
\langle 0 \vert \Phi (w=\zeta) \ . \nfr{bpzstate}
The label ``BPZ'' on the state $\langle \Phi^{\rm BPZ} \vert$ is
necessary in order to avoid confusion with the bra state
$\langle \Phi \vert \equiv
\lim_{\zeta \rightarrow 0} \langle 0 \vert \left( \Phi
(z=\zeta) \right)^{\dagger}$
defined by hermitean conjugation, because this will in general
differ from $\langle \Phi^{\rm BPZ} \vert$. (Another possibility,
preferred by many authors, is to take BPZ conjugation as the defining
map from ket to bra and introduce instead a label $\langle \Phi^{\rm
h.c.} \vert$ on the state defined by hermitean conjugation.)
\section{Composing BPZ and hermitean conjugation}
The composition of BPZ and hermitean conjugation gives a map from
ket to ket
$$
\vert\Phi \rangle\
\longrightarrow\
\left(\langle\Phi^{\rm BPZ}\vert\right)^\dagger\ =\
\vert \Phi^{\rm BPZ}\rangle\ \efr
which acts on the primary conformal fields as follows
$$ \eqalignno{
\Phi_{\Delta,\bar{\Delta}}(z=\zeta,\bar{z}=\bar{\zeta})\ \longrightarrow\
& \left( \Phi_{\Delta,\bar{\Delta}} (w=\zeta,\bar{w}=\bar{\zeta})
\right)^{\dagger}  & \nameali{hermandbpz} \cr
& \ = \
e^{i\epsilon\pi(\Delta-\bar{\Delta})}
\widehat\Phi_{\Delta,\bar{\Delta}}
(z=\zeta^*,\bar{z}=\bar{\zeta}^*) \ . \cr} $$
Notice that for fields with non-integer value of
$\Delta-\bar{\Delta}$,
BPZ and hermitean conjugation do not commute.
However, this is not a
problem for vertex operators describing BRST-invariant on-shell string
states, which satisfy $\Delta=\bar{\Delta}=0$.

The transformation \hermandbpz\ is our educated guess for the map taking
${\cal W}_{\vert \lambda \rangle}$ into
${\cal W}_{\langle \lambda \vert}$, only we will allow the possibility
that some phase factor $\chi$ may appear.
In other words, our ansatz is that if some incoming string
state with definite quantum numbers is created, in the superghost charge
$q$ picture, by the vertex operator
${\cal W}_{\vert\lambda\rangle}^{(q)}$,
$$
\vert \lambda \rangle\ =\ \lim_{z,\bar{z} \rightarrow 0}
{\cal W}_{\vert\lambda\rangle}^{(q)} (z,\bar{z}) \vert 0 \rangle \ ,
\efr
then the vertex operator we have to use in the ``master formula''
\Tmatrix\ to obtain the $T$-matrix element involving the outgoing state
$\langle \lambda \vert$ is given by
$$ {\cal W}_{\langle \lambda \vert}^{(q)} (z=\zeta,\bar{z}=\bar{\zeta})
\ \equiv\ \chi_q \left( {\cal W}_{\vert \lambda \rangle}^{(q)}
(w=\zeta^*,\bar{w} =\bar{\zeta}^*) \right)^{\dagger} \ .
\nfr{guesstwo}
As was emphasized at the beginning of section 4, the operator ${\cal
W}_{\langle \lambda \vert}^{(q)}$, like any vertex operator, creates a
state by acting on the ket vacuum. From the definitions \guesstwo\ and
\bpzstate\ we find this state to be
$$ \lim_{\zeta,\bar{\zeta} \rightarrow 0} {\cal
W}_{\langle \lambda \vert}^{(q)} (z=\zeta,\bar{z}=\bar{\zeta})
\vert 0 \rangle \ = \
\chi_q \vert \lambda^{\rm BPZ} \rangle \ .
\efr
In other words, we obtain the $T$-matrix element involving the bra-state
$\langle \lambda \vert$ by inserting into the Polyakov path integral an
operator creating the state $\chi_q \vert \lambda^{\rm BPZ} \rangle$.
Notice that whereas the state $\vert \lambda \rangle$ always has
$k^0 > 0$, the state $\vert \lambda^{\rm BPZ} \rangle$ has $k^0 < 0$.

Since the combination of BPZ and hermitean conjugation maps
$$ L_0 \rightarrow L_0 \qquad {\rm and} \qquad
Q_{BRST} \rightarrow - Q_{BRST} \ , \efr
and since BPZ conjugation is a world-sheet symmetry on the sphere, it
follows that if the state $\vert \lambda \rangle$ is
a physical on-shell state,
$L_0 \vert \lambda \rangle = Q_{BRST} \vert \lambda \rangle = 0$,
then so is the state $\chi_q \vert \lambda^{\rm BPZ} \rangle$,
regardless of what value we choose for the phase $\chi_q$. It
is less clear that the map \guesstwo\ is also
consistent with the GSO projection, i.e. that $ \vert \lambda \rangle$
satisfies the GSO projection conditions if and only if
$\vert \lambda^{\rm BPZ} \rangle$ does,
because the two states will in general reside in
different sectors of the string theory. An explicit proof in the
framework of a Kawai-Lewellen-Tye (KLT) type
heterotic string model is given in Appendix C.

If we restrict ourselves to BRST invariant on-shell string states, both
${\cal W}_{\vert \lambda \rangle}$ and $\widehat{\cal W}_{\vert
\lambda \rangle}$ are primary conformal fields of dimension zero, and
eq.~\guesstwo\ becomes
$$ {\cal W}^{(q)}_{\langle \lambda \vert} (\zeta,\bar{\zeta}) = \chi_q
\widehat{\cal W}^{(q)}_{\vert \lambda \rangle} (\zeta,\bar{\zeta}) \ .
\nfr{guessthree}
We will now proceed to verify our ansatz \guessthree\
by considering the amplitude
for the string state $\vert \lambda \rangle$ to emit (absorb) a very
soft graviton. We will find that the phase factor $\chi_q$, as
anticipated by our notation, depends only
on the choice of picture. In particular, if we restrict ourselves to the
pictures $q=-1$ and $q=-1/2$, the phase factor $\chi_q$ depends only on
whether the string state is a space-time boson or a space-time
fermion. At the same time we will be able to determine the correct
overall normalization of the vertex operators to be used in the formula
\Tmatrix\ for the $T$-matrix element.
\chapter{Normalization of vertex operators}
In this section we consider the computation of the tree-amplitude for
some given on-shell string state to absorb or emit a very soft graviton.
We perform the
analysis for a generic four dimensional heterotic string theory
where the graviton vertex operator has the form of eq.~\gravvert, but
the argument can be readily applied to
other string models.

We first discuss the case of space-time bosonic states and then
the case of the space-time fermionic states.
\section{Normalization of space-time bosonic vertex operators}
We first recall what is the situation in field theory.
Consider a basis of propagating bosonic particle states with momentum
$p$, labelled by an index $N$, in terms of which the
propagator assumes the diagonal form $P_{MN}/(p^2+ m_N^2)$ where
$P_{MN} = + \delta_{M,N}$ for physical states and
$P_{MN} = -\delta_{M,N}$ for possible
negative norm states. For example, for a photon with space-time vector
index $M=\mu$ we have $P_{MN} = \eta_{\mu \nu}$.

The tree-level $T$-matrix element for such a
particle to emit (absorb) a graviton contains a universal term which, in
the limit where the graviton momentum is zero, assumes the form
$$
\eqalignno{-2\kappa \, \epsilon \cdot p  \ \bar\epsilon \cdot p
\ P_{MN}
&=\ - 4 {\kappa\over \alpha^\prime} \, \epsilon \cdot k \
\bar\epsilon \cdot k  \ P_{MN} &\nameali{rightres}\cr
&=\ - C_0 \left(\kappa\over\pi\right)^3
\, \epsilon \cdot k \ \bar\epsilon \cdot k  \ P_{MN} \ , \cr}
$$
where we wrote the graviton polarization on the factorized form
$\epsilon\otimes\bar\epsilon$ and $C_0$ is the overall
normalization constant for string tree amplitudes, given by
eq.~\overallnorm. The behaviour \rightres\ describes the canonical
coupling of gravity to the $p_{\mu} p_{\nu}$-part of the energy-momentum
tensor of the propagating particle.

The sign of the amplitude \rightres\ obviously depends on the sign
convention for the graviton field $h_{\mu \nu}$.
Eq.~\rightres\ corresponds to the
expansion
$$ g_{\mu \nu} = \eta_{\mu \nu} - 2 \kappa \left( h_{\mu \nu} + \lambda
\eta_{\mu \nu} h^{\sigma}_{\ \sigma} \right) + {\cal O} (h^2)
\nfr{gravfield}
regardless of the coefficient $\lambda$ chosen for the trace term.

The sign chosen for the graviton vertex operator \gravvert\
is in agreement with
this convention, as one may check by computing the 3-graviton tree
amplitude from eqs.~\gravvert\ and \Tmatrix\ and comparing with
eq.~\rightres\ in the case where the state $\vert M \rangle$ is itself
a graviton.

Consider now computing the universal part \rightres\ of the graviton
absorption amplitude at genus zero in string theory. We consider a
complete set of space-time bosonic string states $\vert N,k \rangle$,
labelled by $N$, built from the superghost vacuum $\vert q=-1\rangle$,
satisfying $b_0 = \bar{b}_0 = 0$ and having definite momentum $k$.
We may think of $N$ as specifying physical quantities such as
helicity, charges and family labels.

The $T$-matrix element for the process ``$N + {\rm graviton \ } \rightarrow
M$'' is given by
$$
T^0(M,k \vert {\rm graviton}; N,k)\ =\ -C_0
\vev{ {\cal W}_{\langle M,k \vert}^{(-1)}(z_1,\bar{z}_1)
{\cal W}_{\vert {\rm grav} \rangle}^{(0)}(z,\bar{z})
{\cal W}_{\vert N,k \rangle}^{(-1)}(z_2,\bar{z}_2) } \ ,
\nfr{mgravn}
where we have to use the graviton vertex operator in the superghost
charge $(0)$ picture, given by eq.~\gravverttwo ,
and the states $\vert N,k \rangle$ and $\langle
M,k
\vert$ are now assumed to be physical, so that ${\cal W}^{(-1)}_{\vert
N,k
\rangle}$ and ${\cal W}_{\langle M,k \vert}^{(-1)}$ are primary conformal
fields of dimension zero.

By projective invariance on the sphere we can fix
$z_1=\infty$, $z=1$ and $z_2=0$; and since the
${\cal W}_{\langle M,k \vert}^{(-1)}$
vertex operator is assumed to have conformal dimension zero
we can evaluate it in the coordinate system $(w)$, where
$w=1/z$, without introducing any transformation factor.
In so doing we just undo the BPZ transformation in the definition
eq.~\guesstwo\ of the operator ${\cal W}_{\langle M,k \vert}^{(-1)}$
and obtain
$$ \langle 0 \vert {\cal W}^{(-1)}_{\langle M,k \vert} (w=\bar{w}=0) \ = \
\chi_{-1} \ \langle 0 \vert \left( {\cal W}_{\vert M,k
\rangle}^{(-1)} (z=\bar{z}=0) \right)^{\dagger} \ = \ \chi_{-1} \
\langle M,k \vert \ .
\nfr{fivefour}
Accordingly eq.~\mgravn\ becomes
$$
T^0  (M,k \vert {\rm graviton}; N,k) \ = \ -\chi_{-1} \ C_0
\left({\kappa\over\pi}\right) \vev{ M,k \vert c(1) \bar{c}(1)
\bar\epsilon\cdot\bar\del X (1)
\, \epsilon\cdot\del X(1) \vert N,k }  \ . \nfr{mgravnthree}
Here we may expand the fields $c$, $\bar{c}$, $\partial X$ and
$\bar{\partial} X$ in oscillators. Only modes with $L_0 = \bar{L}_0=0$ can
contribute to the ``universal'' part \rightres\ of the amplitude. This
is because this part of the amplitude, like that of a freely propagating
string state, conserves $L_0(X^{\mu})$, $\bar{L}_0 (X^{\mu})$, $L_0(b,c)$
and $\bar{L}_0 (\bar{b},\bar{c})$. We may imagine the basis
$\vert N,k \rangle$ of string states to diagonalize all these operators.
Then for $n \neq 0$ we may write e.g.
$$ \alpha_n^{\mu} = - {1 \over n} \left[ L_0 (X^{\mu}) ,
\alpha_n^{\mu} \right]
\efr
and this vanishes between the states $\langle M,k \vert$ and $\vert N,k
\rangle$ since by assumption they have the same value of $L_0 (X^{\mu})$.

We are thus left with
$$ T^0 ( M,k \vert {\rm graviton}; N,k ) = -
\chi_{-1} \ C_0 \left({\kappa \over
\pi}\right) \ \epsilon \cdot k \ \bar{\epsilon} \cdot k \
\vev{ M,k \vert \bar{c}_0 c_0 \vert N,k } + \ldots \ , \nfr{mgravntwo}
where ``$+ \ldots$'' denotes possible other terms in the amplitude with a
different kinematical structure than the universal part \rightres .
By eq.~\ghosttwo\ the matrix $\vev{ M,k \vert \bar{c}_0 c_0 \vert N,k }$ is
manifestly hermitean and by an appropriate choice of basis it may be
diagonalized such that
$$ \vev{ M,k \vert \bar{c}_0 c_0 \vert N,k } = \left(
{\cal N}_{\vert M,k\rangle}^{\rm bos}\right)^*
{\cal N}_{\vert N,k \rangle}^{\rm bos} \  P_{MN} \ ,
\nfr{statenorm}
where either $P_{MN} = 0$ (so that the state does not propagate) or
$|P_{MN}| = \delta_{M,N}$. Our conventions \ghostone\ imply that
$P_{MN}=+\delta_{M,N}$ for all physical external states but
$-\delta_{M,N}$ for
negative norm states (such as the ``timelike'' photon). The factor
${\cal N}_{\vert N,k \rangle}^{\rm bos}$
specifies the overall normalization of the
state $\vert N,k \rangle$. By inserting
eq.~\statenorm\ into eq.~\mgravntwo\ we obtain finally the correct
result \rightres\ {\it if} we take the phase factor introduced in
eq.~\guesstwo\ to be $\chi_{-1}=1$ and choose the normalization constant
to be the same for all states,
${\cal N}_{\vert M,k\rangle}^{\rm bos}={\cal N}_{\vert N,k\rangle}^{\rm
bos}$, given by
$$\left| {\cal N}_{\vert N,k \rangle}^{\rm bos} \right|
= {\kappa \over \pi} \ . \efr
In summary,
$$ {\cal W}^{(-1)}_{\langle N,k \vert} (z,\bar{z}) = + \widehat{\cal
W}^{(-1)}_{\vert N,k
\rangle} (z,\bar{z}) \qquad {\rm for \ physical \ spacetime \ bosons} \ ,
\nfr{inoutboson}
and the proper normalization of the state $\vert N,k \rangle$ is given by
$$ \langle M,k \vert \bar{c}_0 c_0 \vert N,k \rangle = \left( {\kappa \over
\pi } \right)^2 P_{MN} \ . \nfr{statenormtwo}
Since by definition
$ \vert N,k \rangle = \lim_{\zeta,\bar{\zeta} \rightarrow 0} {\cal
W}_{\vert N,k \rangle} (\zeta,\bar{\zeta}) \vert 0 \rangle $,
eq.~\statenormtwo\ specifies the normalization of the vertex operator up
to a complex phase factor. If the vertex operator
${\cal W}_{\vert N,k \rangle}^{(-1)}$ is complex, i.e. not proportional to
$\widehat{\cal W}_{\vert N,-k \rangle}^{(-1)}$,~\note{Notice that if
${\cal W}_{\vert N,k \rangle}^{(-1)}$ is proportional to $\exp(ik\cdot X)$
then $\widehat{\cal W}_{\vert N,k \rangle}^{(-1)}$ is proportional to
$\exp(-ik\cdot X)$.} there is probably no
fundamental reason to prefer any specific value of the overall complex
phase factor, just as in field theory the phase of a complex field is
an unphysical degree of freedom. If, on the other hand,
${\cal W}_{\vert N,k \rangle}^{(-1)}$ {\it is} proportional to
$\widehat{\cal W}_{\vert N,-k \rangle}^{(-1)}$ it becomes natural to
impose a reality condition, which we can take to be
$$ {\cal
W}_{\vert N,k \rangle}^{(-1)} = + \widehat{\cal
W}_{\vert N,-k \rangle}^{(-1)} \qquad {\rm or} \qquad
{\cal V}_{\vert N,k \rangle}^{(-1)} = - \widehat{\cal
V}_{\vert N,-k \rangle}^{(-1)} \efr
in agreement with the choice made for the graviton vertex operator
\gravvert . This implies that ${\cal W}_{\langle N,k \vert}^{(-1)} =
{\cal W}_{\vert N,-k \rangle}^{(-1)}$. Even in this case there remains a
choice a sign for the vertex operator. This is completely dependent on
convention, just like the sign of the graviton field in the expansion
\gravfield .
\section{Normalization of space-time fermionic vertex operators}
We now consider the case of space-time fermions. The field theory
description is now more complicated than in the case of space-time
bosons, since the graviton field should be described in terms of the
vierbein, $e^{\mu}_m$. The canonical coupling to gravity of a Dirac
fermion, labelled by an index $N$, is given by the action
$$ \int {\rm d}^4 x\ e \ \overline{\psi}_M \left\{ \gamma^m e^{\mu}_m
\partial_{\mu} + m  \right\} \psi_N \ P^{MN} \ , \efr
where we ignore the spin-connection terms which all involve derivatives
of the vierbein and thus give rise to terms in the
fermion-fermion-graviton amplitude proportional to the graviton
momentum. When expanding $e^{\mu}_m$ around the flat background we can
ignore the deviation of $e=\det \{ e^{\mu}_m \}$ from unity since this
gives rise only to terms proportional to the trace of the graviton
field.
One obtains the following expression, analogous to eq.~\rightres\ for
the universal part of the fermion-fermion-graviton $T$-matrix element at
tree level:
$$ -i \kappa \overline{u}(\vec{p},\eta)
\gamma^{\nu} p^{\mu} u (\vec{p},\eta) \epsilon_{\nu} \bar{\epsilon}_{\mu}
\ P_{MN} \ ,
\nfr{fermrightres}
where, by virtue of the Gordon identity
$$\overline{u}(\vec{p},\eta) \gamma^{\nu} u(\vec{p},\eta') = -2i p^{\nu}
\delta_{\eta,\eta'} \ ,
\nfr{qftgordon}
we recover the bosonic result \rightres , as dictated by the
principle of equivalence.

In the string theory analysis we again consider a complete set of
states $\vert N,k \rangle$, labelled by $N$,
now built from the superghost vacuum $\vert q=-1/2 \rangle$,
again satisfying $b_0 = \bar{b}_0 = 0$ and having a definite momentum
$k$.

We may now proceed exactly as in section 5.1, only now we have to use
the superghost charge $(-1)$ version of the graviton vertex operator,
given by eq.~\gravvert.
In the limit of vanishing graviton momentum we obtain
$$
\eqalignno{ T^0(M,k \vert {\rm graviton} ; N,k)\  &= \
-C_0 \ \vev{ {\cal W}_{\langle M,k \vert}^{(-1/2)}(z_1,\bar{z}_1)
{\cal W}_{\vert {\rm grav} \rangle}^{(-1)}(z,\bar{z})
{\cal W}_{\vert N,k \rangle}^{(-1/2)}(z_2,\bar{z}_2) } \cr
& = \ - \chi_{-1/2} \ C_0 \ \langle M,k \vert
{\cal W}_{\vert {\rm grav} \rangle}^{(-1)}(1) \vert N,k \rangle \ .
&\nameali{vmandvnone} \cr } $$
As in the bosonic case only zero-mode operators contribute to the part
of the amplitude in which we are interested, so that
$$ T^0(M,k \vert {\rm graviton} ; N,k)  \ = \
\chi_{-1/2} \ C_0 {\kappa \over \pi } \ \bar{\epsilon} \cdot k \
\epsilon_{\nu} \langle M,k \vert
\bar{c}_0 c_0 \psi_0^{\nu} \delta(\gamma_0)
\vert N,k \rangle + \ldots  \ .
\nfr{vmandvn}
Here we may recognize the form \fermrightres\ of the result obtained in
field theory, since the zero mode $\psi_0^{\nu}$ of the operator field
$\psi^{\nu}$ furnishes a representation of the Clifford algebra, and so
is completely analogous to the gamma matrix $\gamma^{\nu}$ appearing in
the expression \fermrightres .

The matrix $\langle M,k \vert \bar{c}_0 c_0 \psi_0^{\nu} \delta(\gamma_0)
\vert N,k \rangle$ transforms as a space-time vector and therefore has to
be proportional to the momentum $k^{\nu}$. Since $\psi_0^{\nu}$ and
$\delta(\gamma_0)$ anti-commute it is manifestly
anti-hermitean (q.v. eq.~\ghosttwo )
and by choosing an appropriate basis it can be
diagonalized such that
$$\langle M,k \vert \bar{c}_0 c_0 \psi_0^{\nu} \delta(\gamma_0)
\vert N,k \rangle = i Y k^{\nu} \left(
{\cal N}_{\vert M,k \rangle}^{\rm ferm} \right)^*
{\cal N}_{\vert N,k \rangle}^{\rm ferm} \ P_{MN} \ .
\nfr{fermnorm}
In section 7 we will explicitly derive this formula in the context of a
KLT heterotic string model. It is quite analogous to eq.~\statenorm .
The factor of $i$ reflects the fact that the matrix on the left-hand
side is anti-hermitean and (as we shall see in section 7)
the constant factor $Y=\pm 1$ depends on the conventions chosen for the
spin fields.
Finally, $P_{MN} = + \delta_{M,N}$ for physical states, as always.
Like in the bosonic case the factor
${\cal N}_{\vert N,k \rangle}^{\rm ferm}$ specifies the normalization of
the string state $\vert N,k \rangle$.
If we insert eq.~\fermnorm\ into eq.~\vmandvn\ we finally obtain
$$\eqalignno{
& T^0 ( M,k \vert {\rm graviton}; N,k ) & \nameali{fivetwenty} \cr
& \qquad \qquad = \
 Y \ i \ \chi_{-1/2} \ C_0 \left(
{\cal N}_{\vert M,k \rangle}^{\rm ferm} \right)^*
{\cal N}_{\vert N,k \rangle}^{\rm ferm}
\left( \kappa \over \pi \right)
\bar{\epsilon} \cdot k \ \epsilon \cdot k \ P_{MN} + \ldots \ , \cr} $$
which reproduces the right result \rightres\ assuming we choose
$$ \chi_{-1/2} = i Y \qquad {\rm for \ spacetime \ fermions}
\nfr{ysign}
and fix the normalization of the states in the same universal way as for
the bosons,
${\cal N}_{\vert M,k \rangle}^{\rm ferm}={\cal N}_{\vert N,k \rangle}^{\rm
ferm}$, and
$$ \left| {\cal N}_{\vert N,k \rangle}^{\rm
ferm} \right| = \left| {\cal N}_{\vert N,k \rangle}^{\rm
bos} \right| = {\kappa \over \pi} \ . \nfr{uninorm}
In summary
$$
{\cal W}_{\langle N,k \vert}^{(-1/2)} (z,\bar{z})\ =\ (i Y)\,
\widehat{\cal W}_{\vert N,k \rangle}^{(-1/2)}(z,\bar{z}) \qquad
{\rm for \ physical \ spacetime \ fermions,}
\nfr{fermap}
and the proper normalization of the string state is given by
$$\langle M,k \vert \bar{c}_0 c_0 \psi_0^{\nu} \delta(\gamma_0)
\vert N,k \rangle = i Y k^{\nu} \left( {\kappa \over \pi } \right)^2
\ P_{MN} \ .
\nfr{fermnormtwo}
Since the PCO \PCO\ is an anti-hermitean operator which satisfies Bose
statistics, eqs.~\inoutboson\ and \fermap\ can be generalized to the
superghost charge $q$ picture as follows
$$ {\cal W}_{\langle N,k \vert}^{(q)} (z,\bar{z}) \ = \ \chi_{q} \
\widehat{\cal W}_{\vert N,k \rangle}^{(q)} (z,\bar{z}) \ = \
(-1)^{q+1} \ \widehat{\cal W}_{\vert N,k \rangle}^{(q)} (z,\bar{z})
 \ . \nfr{inoutq}
For pictures of half-integer $q$ (i.e. pictures describing space-time
fermions) the phase factor $(-1)^{q+1}$ involves a choice of sign, which
is parametrized by $Y$ according to eq.~\ysign , i.e. $(-1)^{1/2} = iY$.
\chapter{Space-Time hermiticity}
An important check on the correctness of our expressions
\inoutboson\ and \fermap\ for ${\cal W}_{\langle N,k \vert}^{(q)}$
is provided by the requirement that
the $T$-matrix element obtained from eq.~\Tmatrix\ has the right
hermiticity properties.

Unitarity requires that the tree-level $T$-matrix element is real except
when the momentum flowing in some intermediate channel happens to be on
the mass-shell corresponding to some physical state in the theory.
In field theory the imaginary part appears as a result of the
$i\epsilon$-prescription present in the propagator that happens to be
on-shell. In string theory it appears as a result of some divergency in
the integral over the Koba-Nielsen (KN) variables that has to be treated in a
way consistent with the $i\epsilon$-prescription in field
theory~[\Ref{Hoker},\Ref{Berera},\Ref{Weisberger}].

What we can rather easily show is that as long as the integrals over the
KN variables are convergent the expressions
\inoutboson\ and \fermap\ lead to a hermitean $T$-matrix at tree level.

At genus zero the formula \Tmatrix\ can be rewritten as
$$
\eqalignno{ & T^0
(\lambda_1; \dots ; \lambda_{N_{out}} \vert
\lambda_{N_{out}+1};\dots;\lambda_{N_{out} + N_{in}} ) \ =
& \nameali{Ttree} \cr
&\qquad  - C_0 \ \int  \left( \prod_{i=4}^{N_{tot}}
 \di^2 z_i \right) \
\vev{ \bar{c}(\bar{z}_1) \bar{c}(\bar{z}_2) \bar{c}(\bar{z}_3)
c(z_1) c(z_2) c(z_3)\ \times  \cr
&\qquad\quad \left(\prod_{A=1}^{N_B+N_{FP}-2}
\Pi (w_A)\right) \, {\cal V}_{\langle \lambda_1 \vert} (z_1,\bar{z}_1)
\dots {\cal V}_{\vert \lambda_{N_{\rm tot}} \rangle}
(z_{N_{\rm tot}},\bar{z}_{N_{\rm tot}})}  \ . \cr}
$$
The $T$-matrix is hermitean if and only if the quantity \Ttree\ equals
$$
\eqalignno{ & \left[ T^0 ( \lambda_{N_{out} + N_{in}}; \ldots ;
\lambda_{N_{out}+1} \vert \lambda_{N_{out}} ; \ldots ;
\lambda_1 ) \right]^*\ = & \nameali{Ttreeherm} \cr
&\qquad + C_0 \ \int \left( \prod_{i=4}^{N_{\rm tot}}
\di^2 z_i^* \right) \
\vev{ \left( {\cal V}_{\vert \lambda_1 \rangle} (z_1,\bar{z}_1)
\right)^{\dagger} \ldots \left(
{\cal V}_{\langle \lambda_{N_{\rm tot}} \vert}
(z_{N_{\rm tot}},\bar{z}_{N_{\rm tot}}) \right)^{\dagger}\ \times \cr
&\qquad\ \left( \Pi (w_{N_B + N_{FP}-2}) \right)^{\dagger} \ldots
\left( \Pi(w_1) \right)^{\dagger} \ \left( c(z_3) \right)^{\dagger}
\dots \left( \bar{c}(\bar{z}_1) \right)^{\dagger} }  \ , \cr }
$$
where we used eq.~\ghostherm.

In terms of the vertex operators ${\cal V}$ (where the $c\bar{c}$ factor
present in ${\cal W}$ has been removed, q.v. eq.~\wvvect) the relations
\inoutboson\ and \fermap\ acquire an extra
minus sign (because $c\bar{c}$ is an anti-hermitean operator):
$$\eqalignno{ {\cal V}^{(-1)}_{\langle \lambda \vert} (z,\bar{z}) & = \
- \widehat{\cal V}_{\vert \lambda \rangle}^{(-1)} (z,\bar{z}) &
\nameali{inoutv} \cr
{\cal V}^{(-1/2)}_{\langle \lambda \vert} (z,\bar{z}) & = \
- i Y \ \widehat{\cal V}_{\vert \lambda \rangle}^{(-1/2)} (z,\bar{z})
\ , \cr } $$
which, by taking the hermitean conjugate, leads to the inverse relations
$$\eqalignno{ {\cal V}^{(-1)}_{\vert \lambda \rangle} (z,\bar{z}) & = \
- \widehat{\cal V}_{\langle \lambda \vert}^{(-1)} (z,\bar{z}) &
\nameali{outinv} \cr
{\cal V}^{(-1/2)}_{\vert \lambda \rangle} (z,\bar{z}) & = \
- i Y \ \widehat{\cal V}_{\langle \lambda \vert}^{(-1/2)} (z,\bar{z})
\ . \cr } $$
Since the operators ${\cal V}_{\vert \lambda \rangle}$ and
${\cal V}_{\langle \lambda \vert}$ have conformal dimensions
$\Delta=\bar{\Delta} = 1$ we find for $i=1,\ldots,N_{out}$:
$$\eqalignno{  \left( {\cal V}_{\vert \lambda_i \rangle}
(z_i,\bar{z}_i) \right)^{\dagger} \ &= \ \left( {1 \over z_i^*} {1 \over
\bar{z}_i^*} \right)^2 \widehat{\cal V}_{\vert \lambda_i \rangle} \left(
{1 \over z_i^*}, {1 \over \bar{z}_i^*} \right) & \nameali{voptrans} \cr
& = ( {\rm phase \ factor} ) \times \left( {1 \over z_i^*} {1 \over
\bar{z}_i^*} \right)^2 \times {\cal V}_{\langle \lambda_i \vert}
\left( {1 \over z_i^*}, {1 \over \bar{z}_i^*} \right) \ , \cr }
$$
where the phase factor we pick up is minus one for space-time bosons and
$iY$ for space-time fermions. By eqs.~\outinv\ we pick up exactly the
same phase factor from vertex operators of the type ${\cal V}_{\langle
\lambda \vert}$. This amounts to an overall sign $(-1)^{N_B + N_{FP}}$,
$N_{FP}$ being the number of space-time fermion pairs and $N_B$
the number of space-time bosons. This sign exactly cancels the sign
produced by the $N_B + N_{FP}-2$ PCOs, which are anti-hermitean.
Finally, reordering the ghost factors in \Ttreeherm\ in accordance with
eq.~\Ttree , we obtain a minus sign cancelling the one that was
introduced by using eq.~\ghostherm.

Since the transformation factors $(z_i^*)^{-2} (\bar{z}_i^*)^{-2}$
appearing in eq.~\voptrans\ either cancels a similar one coming from the
ghost operators (for $i=1,2,3$), or is just the required jacobian to
transform $\di^2 z_i$ into $\di^2 \zeta_i$ where $\zeta_i = 1/z_i^*$ ($i
\geq 4$), we finally recover eq.~\Ttree\ multiplied by a phase factor
that, at the end, is just plus one.
This concludes the proof that our relation between
${\cal W}_{\langle \lambda \vert}^{(q)}$ and
${\cal W}_{\vert \lambda \rangle}^{(q)}$ leads to a hermitean
$T$-matrix at tree level away from the resonances.
\chapter{An explicit example}
In this section we provide an explicit example of the map \inoutq\
in the context of four-dimensional heterotic string models of the
Kawai-Lewellen-Tye (KLT) type [\Ref{KLT},\Ref{Anto}], where the internal
degrees of freedom are described by 22 left-moving and 9 right-moving
free complex fermions. We bosonize all these fermions (as well as the
four Majorana fermions $\psi^{\mu}$), using the explicit prescription
for bosonization in Minkowski space-time proposed in ref.~[\Ref{mink}].

In this formulation any state of the conformal field theory
(excluding the reparametrization ghosts) can be obtained by means of
non-zero mode creation operators from the generic ground state
which is specified by the
space-time momentum $k$, the ``momentum'' $J_0^{(L)} = {\Bbb A}_L$
of the 33 bosons $\Phi_{(L)}$ introduced by the bosonization,
and the superghost charge
$J_0^{(34)}=q={\Bbb A}_{34}$ which is (minus) the ``momentum''
of the field $\phi \equiv \Phi_{(34)}$ that is introduced when ``bosonizing''
the superghosts. Since $[ J_0^{(L)} , \Phi_{(K)} ] = \delta_K^{\ L}$,
the operator creating such a ground state
from the conformal vacuum is
$$ S_{\Bbb A} (z,\bar{z}) \ e^{i k \cdot X(z,\bar{z})} \ ,
\nfr{groundst}
where
$$ S_{\Bbb A} (z,\bar{z}) \equiv \prod_{L=1}^{34} e^{{\Bbb A}_L
\Phi_{(L)} (z,\bar{z})} \left( C_{(L)} \right)^{{\Bbb A}_L} \ ,
\nfr{spinfield}
is a spin field operator and $C_{(L)}$ is a cocycle factor, see
ref.~[\Ref{mink}] for details.

The range of values allowed for the ${\Bbb A}_L$ depends on the details
of the KLT model we happen to consider, see
refs.~[\Ref{KLT},\Ref{ammedm}]. We assume the level-matching condition
$L_0 - \bar{L}_0 = 0$ to be satisfied.

The hermitean conjugate of the operator $S_{\Bbb A} (z,\bar{z})$ can be
computed using the hermiticity properties of the various fields, as
outlined in Appendix B (see also ref.~[\Ref{mink}]). One finds
$$ \widehat{S}_{\Bbb A} (z,\bar{z}) = \left( \sigma_1^{(33)} {\Bbb
C}^{-1} \right)_{\Bbb A}^{\ \, \Bbb B} \ S_{\Bbb B} (z,\bar{z}) \ ,
\nfr{hermground}
where
$$ (\sigma_1^{(33)})_{{\Bbb A} {\Bbb B}} =
\left( \prod_{L=1}^{32} \delta_{{\Bbb A}_L,{\Bbb B}_L}\right) \ \delta_{{\Bbb
A}_{33} + {\Bbb B}_{33}, 0} \ \delta_{{\Bbb A}_{34},{\Bbb B}_{34}} \efr
and ${\Bbb C}^{-1}$ is the inverse of the ``charge conjugation matrix''
$$ {\Bbb C}_{{\Bbb A} {\Bbb B}} =
\left( \prod_{L=1}^{33} \delta_{{\Bbb A}_L + {\Bbb B}_L,0}  \right) \
\delta_{{\Bbb A}_{34},{\Bbb B}_{34}} \ e^{ i \pi {\Bbb A} \cdot Y
\cdot {\Bbb B}}
\nfr{chargeconj}
defined in terms of the $34 \times 34$ cocycle matrix $Y_{KL}$ (see
refs.~[\Ref{mink},\Ref{ammedm}]).

The example we want to study is that of a physical
space-time fermion described by a ground state. To obtain a
BRST-invariant state one has to consider a vertex operator
which involves a linear combination of spin fields,
$$ {\cal V}^{(-1/2)}_{\vert {\Bbb V}, k \rangle} (z,\bar{z}) \ = \
{\kappa \over \pi} \ {\Bbb V}_{(-{1 \over 2})}^{\Bbb A} (k) \, S_{\Bbb A}
(z,\bar{z}) \ e^{i k \cdot X(z,\bar{z})} \ , \nfr{groundtwo}
where the spinor ${\Bbb V}^{\Bbb A}_{(-{1 \over 2})} (k)$ has superghost
charge $-1/2$, i.e. is proportional to
$\delta_{{\Bbb A}_{34},-1/2}$, and satisfies a Dirac equation which can
be obtained from the requirement that the $3/2$-order pole in the
operator product expansion (OPE) of the supercurrent $T_F^{[X,\psi]}$
with the operator \groundtwo\ vanishes.
If we define the gamma matrices by the OPE
$$ \psi^{\mu} (z) S_{\Bbb A} (w,\bar{w}) \eqope {1 \over \sqrt{2}}
\left( {\bfmath \Gamma}^{\mu} \right)_{\Bbb A}^{\ \Bbb B} S_{\Bbb B}
(w,\bar{w}) {1 \over \sqrt{z-w}} + \ldots \ , \nfr{gammadef}
the Dirac equation assumes the matrix form
$$ ({\Bbb V}_{(-{1 \over 2})} (k))^T \ {\Bbb D} \, (k) = 0 \qquad {\rm or}
\qquad
\left( {\Bbb D}\, (k) \right)^T {\Bbb V}_{(-{1 \over 2})} (k) = 0 \ ,
\nfr{Diraceq}
where the Dirac operator is
$$ {\Bbb D}\, (k) = k_{\mu} {\bfmath \Gamma}^{\mu} - {\Bbb M} \ , \efr
${\Bbb M}$ being a mass operator that we do not need to write down
explicitly.

When the vertex operator is written as in eq.~\groundtwo\ we are no
longer free to choose the normalization of the spinor
${\Bbb V}_{(-{1 \over 2})} (k)$. It should be fixed in accordance
with eq.~\fermnormtwo .
In the next subsection we will explicitly verify that the
correct normalization is
$$ ({\Bbb V}_{(-{1 \over 2})} (k))^{\dagger} \
{\Bbb V}_{(-{1 \over 2})} (k) = \sqrt{2} \ |k^0| \ ,
\nfr{stringspinornorm}
which is analogous in structure to eq.~\ftspinornorm .

By using eq.~\hermground\ in the expression \inoutv\ we find the
``outgoing'' vertex operator corresponding to \groundtwo\ to be
$$ {\cal V}^{(-1/2)}_{\langle {\Bbb V}, k \vert} (z,\bar{z}) \ = \ -
\chi_{-1/2} {\kappa \over \pi} \left( {\Bbb V}^{{\Bbb A}}_{(-{1 \over 2})}
(k)\right)^* \left( \sigma_1^{(33)}
{\Bbb C}^{-1} \right)_{\Bbb A}^{\ \, {\Bbb B}} S_{\Bbb B}
(z,\bar{z}) \ e^{-i k \cdot X(z,\bar{z})} \ , \nfr{groundout}
where $\chi_{-1/2} = i Y$.
\section{A Sample Computation.}
We will now explicitly compute the amplitude for a space-time fermion
described by the vertex operator \groundtwo\ to absorb a zero-momentum
graviton. In particular we will obtain the relation \fivetwenty\ and show
how the sign $Y$ appearing in this formula is related to the choice of
cocycles.

Inserting eqs.~\groundtwo , \groundout\ and \gravvert\ into
eq.~\vmandvnone\ we obtain:
$$\eqalignno{T^0 ({\Bbb V},k \vert {\rm graviton} ; {\Bbb V}, k )
& = \ - C_0 \ \langle {\cal W}_{\langle {\Bbb V},k \vert}^{(-1/2)}
(z_1,\bar{z}_1) \ {\cal W}_{\vert {\rm grav} \rangle}^{(-1)}
(z,\bar{z}) {\cal W}_{\vert {\Bbb V},k \rangle}^{(-1/2)}
(z_2,\bar{z}_2) \rangle &\nameali{ampl} \cr
& =\ i \chi_{-1/2} C_0 \left( {\kappa \over \pi} \right)^3 \left( {\Bbb
V}^{\Bbb A}_{(-{1 \over 2})} (k) \right)^*
\left( \sigma_1^{(33)} {\Bbb C}^{-1} \right)_{\Bbb A}^{\ \, {\Bbb B}}
{\Bbb V}^{\Bbb C}_{(-{1 \over 2})} (k) \ \epsilon_{\mu}\ \times \cr
& \quad \langle S_{\Bbb B}(z_1,\bar{z}_1) \psi^{\mu}(z)
e^{- \Phi_{(34)} (z)} (C_{(34)})^{-1} S_{\Bbb C} (z_2,\bar{z}_2) \rangle
\ \times\cr
& \quad \langle \bar{\epsilon} \cdot \bar{\del} X(\bar{z}) e^{-i k \cdot
X(z_1,\bar{z}_1)} e^{i k \cdot X (z_2,\bar{z}_2)} \rangle \
\langle \bar{c} (\bar{z}_1) \bar{c} (\bar{z}) \bar{c} (\bar{z}_2)
c (z_1) c (z) c (z_2) \rangle \ . \cr }
$$
By explicit computation one finds
$$\eqalignno{ & \langle e^{-\Phi_{(34)} (z) } (C_{(34)})^{-1} \psi^{\mu}
(z) S_{\Bbb B}(z_1,\bar{z}_1) S_{\Bbb C} (z_2,\bar{z}_2) \rangle\ = &
\nameali{corr} \cr
&\qquad \ {1 \over \sqrt{2}} \left( {\bfmath \Gamma}^{\mu} \ {\Bbb C}_{(-1)}
\right)_{{\Bbb B}\, {\Bbb C}} { z_1 - z_2 \over (z-z_1)(z-z_2)} | z_1 -
z_2 |^{-2(2+m^2)} \ , \cr }
$$
where $m$ is the mass of the space-time fermion, $k^2 + m^2 = 0$, and we
introduced another family of ``charge conjugation matrices'' by
$$ \left( {\Bbb C}_{(q)} \right)_{{\Bbb A} {\Bbb B}} =  \left(
\prod_{L=1}^{33} \delta_{{\Bbb A}_L + {\Bbb B}_L,0} \right) \
\delta_{{\Bbb A}_{34} +
{\Bbb B}_{34} + q + 2, 0 } \ e^{ i \pi {\Bbb A} \cdot Y \cdot {\Bbb B}}
\efr
for any value of $q \in {\Bbb Z}$.

Similarly one finds
$$\eqalignno{& \langle \bar{\epsilon} \cdot \bar\del X (\bar{z}) e^{-i k
\cdot X (z_1,\bar{z}_1)} e^{i k \cdot X (z_2,\bar{z}_2)} \rangle \ = \
i \bar{\epsilon} \cdot k {\bar{z}_1 - \bar{z}_2 \over (\bar{z} -
\bar{z}_1) (\bar{z} - \bar{z}_2) } | z_1 - z_2 |^{-2 k^2}\cr
& \langle \ \vert c(z_1) c(z) c(z_2) \vert^2 \ \rangle \ = \
\vert (z_1 - z) ( z- z_2) (z_1 - z_2) \vert ^2 \ . & \nameali{corrr} \cr }
$$
Substituting \corr\ and \corrr\ into eq.~\ampl\ we obtain
$$\eqalignno{ & T^0 ({\Bbb V}, k \vert {\rm graviton}; {\Bbb V}, k )\ = &
\nameali{ampltwo} \cr
&\  \chi_{-1/2} C_0 \left( {\kappa \over \pi} \right)^3 {1 \over
\sqrt{2} } \ \bar{\epsilon} \cdot k \
\epsilon_{\mu} \left( \left( {\Bbb V}_{(-{1 \over 2})} (k) \right)^{\dagger}
\sigma_1^{(33)} {\Bbb C}^{-1} {\bfmath \Gamma}^{\mu} {\Bbb C}_{(-1)}
{\Bbb V}_{(-{1 \over 2})}(k) \right) \ . \cr } $$
One may show that
$$ \left( {\bfmath \Gamma}^{\mu} {\Bbb
C}_{(-1)} \right)_{{\Bbb A} {\Bbb B}} \ = \
(-1)^{{\Bbb A}_{34}+1/2} \left( {\Bbb C}_{(-1)} {\bfmath \Sigma}
\left( {\bfmath \Gamma}^{\mu} \right)^T \right)_{{\Bbb A} {\Bbb B}} \ ,
\nfr{identityone}
where
$$ {\bfmath \Sigma}_{{\Bbb A} {\Bbb B}} \equiv \left( \prod_{L=1}^{34}
\delta_{{\Bbb A}_L,{\Bbb B}_L} \right) \
\exp \left\{ i \pi \sum_{L=1}^{33} Y_{34,L} {\Bbb B}_L \right\}  \efr
and the sign $(-1)^{{\Bbb A}_{34}+1/2}$ is effectively equal to one,
since the matrices appearing in eq.~\ampltwo\ are sandwiched between
spinors with superghost charge $-1/2$. For the same reason the inverse
charge conjugation matrix ${\Bbb C}^{-1}$ is effectively equal to
$({\Bbb C}_{(-1)})^{-1}$.
Finally it is straightforward to verify that ${\bfmath \Gamma}^0$, as
defined by eq.~\gammadef , may also be written on the form
$$ {\bfmath \Gamma}^0 = i Y_{34,33} {\bfmath \Sigma} \sigma_1^{(33)}
\nfr{gammazero}
and since $\bfmath\Sigma$ and $\sigma_1^{(33)}$ anticommute,
$({\bfmath\Gamma}^0)^T = - {\bfmath\Gamma}^0$.

Inserting eqs.~\identityone\ and \gammazero\ into eq.~\ampltwo\ we
obtain
$$ \eqalignno{
& T^0 ({\Bbb V}, k \vert {\rm graviton}; {\Bbb V}, k )\ = & \numali \cr
& \qquad - i \chi_{-1/2} Y_{34,33} \ C_0 \left( {\kappa \over \pi} \right)^3
{1 \over \sqrt{2} } \ \bar{\epsilon} \cdot k \
\epsilon_{\mu} \left( \left( {\Bbb V}_{(-{1 \over 2})} (k) \right)^{\dagger}
({\bfmath \Gamma}^0)^T ({\bfmath \Gamma}^{\mu})^T
{\Bbb V}_{(-{1 \over 2})} (k) \right) \ . \cr } $$
At this point we may use the Gordon-like identity
$$
\left( {\Bbb V}_{(-{1 \over 2})}(k) \right)^{\dagger}
({\bfmath \Gamma}^0)^T ({\bfmath \Gamma}^{\mu})^T  \
{\Bbb V}_{(-{1 \over 2})}(k) = - \sqrt{2} k^{\mu} \ . \efr
This equation can be proven directly using the Dirac equation \Diraceq ,
but it is easier to note that Lorentz covariance forces the right-hand
side to be proportional to $k^{\mu}$ and then fix
the proportionality constant by setting $\mu = 0$ and
using equation \stringspinornorm. Thus we finally obtain
$$ T^0 ({\Bbb V}, k \vert {\rm graviton}; {\Bbb V}, k ) \ = \
 i \chi_{-1/2} Y_{34,33} \ C_0 \left( {\kappa \over \pi} \right)^3
\bar{\epsilon} \cdot k \
\epsilon \cdot k \ . \efr
This agrees with the correct result \rightres\ provided we choose
$$\chi_{-1/2} = i Y = i Y_{34,33} \efr
and shows that the sign $Y$ appearing in eq.~\fermnorm\ should be
identified with the component $Y_{34,33}$ of the cocycle matrix. At the
same time we have verified the correctness of the normalization
\stringspinornorm\ for the spinor ${\Bbb V}_{-{1 \over 2}} (k)$.
\appendix{Conventions for operators and partition functions.}
In this appendix we summarize our conventions for operator fields,
partition functions and spin structures in the explicit setting of a
Kawai-Lewellen-Tye (KLT) heterotic string model. For more details, see
refs.~[\Ref{mink}] and [\Ref{ammedm}].

{\bf Space-time coordinate field:}
$$\eqalignno{& X^{\mu} (z,\bar{z}) \ = \
q^{\mu} - i k^{\mu} \log z - i k^{\mu} \log \bar{z}
+ i \sum_{n \neq 0} {a_n^{\mu} \over n} z^{-n} + i
\sum_{n \neq 0} {\bar{a}_n^{\mu} \over n} \bar{z}^{-n} &\numali\cr
& X^{\mu} (z,\bar{z}) X^{\nu} (w,\bar{w}) \eqope - \eta^{\mu \nu} \log
(z-w) + {\rm \ c.c.\ } + \ldots &\numali\cr
& \wew{1}_{g-{\rm loop}} = (\det \bar{\del}_0 )^{-D/2} (\det 2\pi {\rm
Im} \tau )^{-D/2} \ , &\numali\cr}
$$
where $\tau$ is the period matrix (as given in ref.~[\Ref{PDV1}])
and the explicit expression for $\det
\bar{\del}_0$ can be found for example in ref.~[\Ref{paolosewing}]. It
is normalized to give plus one in the limit where all loops are pinched.

{\bf Majorana fermion field:}
$$ \psi^{\mu} (z) = \sum_n \psi_n^{\mu} z^{-n-1/2} \qquad \qquad
\{ \psi^{\mu}_n, \psi^{\nu}_m \} = \eta^{\mu \nu} \delta_{n+m,0} \ ,
\efr
where the mode index $n$ is integer (half odd integer) for Ramond
(Neveu-Schwarz) boundary conditions. The anti-commutation relations are
equivalent to the OPE
$$ \psi^{\mu} (z) \psi^{\nu} (w) \eqope \eta^{\mu \nu} {1 \over z-w} +
\ldots \ . \efr
When computing correlation functions we usually bosonize all fermion
fields.

{\bf Bosonized complex fermion:}
$$\eqalignno{ & \phi(z) \phi(w) \eqope \log (z-w) + \ldots &\numali\cr
& \wew{\prod_{i=1}^N e^{q_i \phi(z_i) } }_{g-{\rm loop}} \ = & \numali \cr
&\quad\delta_{\sum_{i=1}^N q_i,0} ( \det \bar{\del}_0 )^{-1/2} \prod_{i < j}
\left(E(z_i,z_j)\right)^{q_i q_j}\ \Teta{\alpha}{\beta}
\left( \sum_{i=1}^N q_i \int^{z_i} {\omega \over 2\pi i} \vert \tau \right)
\ , \cr }
$$
where $\omega_{\mu}$ is normalized to have period $2\pi i
\delta_{\mu,\nu}$
around the cycle $a_{\nu}$ ($\mu,\nu = 1, \ldots, g$), $E(z,w)$ is the
prime form (with short-distance behaviour $E(z,w) = (z-w) + {\cal O}
(z-w)^2$) and we define
$$\eqalignno{ & \Teta{\alpha}{\beta} (z \vert \tau) \ = \
\sum_{r \in {\Bbb Z}^g} \exp \left\{ 2\pi i \left[ {1 \over 2}
\sum_{\mu,\nu=1}^g (r_{\mu} + {1 \over 2} - \alpha_{\mu} ) \tau_{\mu \nu}
(r_{\nu} + {1 \over 2} - \alpha_{\nu} ) \right. \right. \cr
& \qquad\qquad + \left. \left.
\sum_{\mu=1}^g (r_{\mu} + {1 \over 2} -
\alpha_{\mu} ) (z_{\mu} + \beta_{\mu} + {1 \over 2}) \right] \right\} \ .
&\numali \cr } $$

{\bf Superghosts:}

\noindent Our conventions for mode expansions, OPEs and ``bosonization''
of the superghosts $\beta$ and $\gamma$ are the standard ones~[\Ref{FMS}].
We always remain inside the ``little''
algebra, i.e. excluding the zero mode of $\eta$ and $\xi$.
Our convention for the partition function is
$$\eqalignno{ & \wew{ \prod_{i=1}^N e^{q_i \phi(z_i)} }_{g-{\rm loop}}
\ = & \numali \cr
&\quad\delta_{\sum_{i=1}^N q_i - 2g+2,0} \
(\det \bar\del_0 )^{1/2} \ \prod_{i=1}^N \left( \sigma(z_i)
\right)^{-2 q_i}  \ \prod_{i < j} \left( E(z_i,z_j) \right)^{-q_iq_j}
\ \times \cr
&\quad \prod_{\mu=1}^g \left( e^{-2\pi i (1/2+ \beta_{\mu})} \right)\
\left[ \Teta{\alpha}{\beta} \left( - \sum_{j=1}^N q_j \int_{z_0}^{z_j}
{\omega \over 2\pi i} + 2 \Delta^{z_0} \vert \tau \right) \right]^{-1} \
. \cr }
$$
This expression agrees with eq.~(36) of ref.~[\Ref{Verlinde}], except
for the overall sign which differs in two regards: First there is the
phase factor appearing at the beginning of the third line above, which is
chosen in accordance with our definition of the spin structure summation
coefficient given below. Second, the sign of the argument of the theta
function is opposite to that of ref.~[\Ref{Verlinde}], which amounts to
a factor of minus one for odd spin structures. The sign we quote above
for the argument of the theta function
is the one that is obtained when the correlation function is carefully
constructed by sewing [\Ref{PDV2}]. Hence it is the sign consistent
with factorization. Our conventions for the differential $\sigma$ and
the Riemann class $\Delta^{z_0}$ are in accordance with ref.~[\Ref{PDV1}].

{\bf Reparametrization ghosts:}

\noindent Our conventions for reparametrization ghosts follow
ref.~[\Ref{FMS}]. The normalization of the partition function is
the standard one, and the explicit expressions can be found in
refs.~[\Ref{Kaj2},\Ref{Paolo}].
By definition the correlator
$$ \wew{\left| \prod_{I=1}^{3g-3+N_{\rm tot}} (\eta_I \vert b)
\prod_{i=1}^{N_{\rm tot}} c(z_i) \right|^2 } \efr
is positive definite.

{\bf Spin structure summation coefficient:}

\noindent Finally the spin structure summation coefficient in
eq.~\Tmatrix\ is a product of one-loop summation coefficients which are
given in accordance with ref.~[\Ref{ammedm}] by
$$\eqalignno{ & C^{{\bfmath\alpha}_{\mu}}_{{\bfmath \beta}_{\mu}} \ = \
{1 \over \prod_i M_i } \times & \numali \cr
&\qquad \exp \left\{ - 2\pi i \left[ \sum_i (n_i^{\mu} + \delta_{i,0})
( \sum_j k_{ij} m_j^{\mu} + s_i - k_{i0} ) + \sum_i m_i^{\mu} s_i +
{1 \over 2} \right] \right\} \ , \cr } $$
where $\mu = 1, \ldots, g$.
The spin structure $\left[ {}^{\alpha_L}_{\beta_L} \right]$ of the
fermion labelled by $L \in \{1, \ldots, 32\}$ is related to the integers
$m_i^{\mu}$ and $n_i^{\mu}$ through
$$\eqalignno{ \alpha_{L,\mu}  & = \ \sum_i m_i^{\mu} \left( {\bfmath W}_i
\right)_{(L)} & \numali \cr
\beta_{L,\mu}  & = \ \sum_i n_i^{\mu} \left( {\bfmath W}_i \right)_{(L)} \ .
\cr}
$$
For more details, see ref.~[\Ref{ammedm}].

\appendix{Hermiticity properties of operators}
In this appendix we summarize the hermiticity properties of various
primary operators and provide some of the details in the derivation of
eq.~\hermground .

The hermitean conjugate field $\widehat{\Phi}_{\Delta}$ of a primary
conformal field $\Phi_{\Delta}$ of dimension $\Delta$ is defined by
eq.~\hermconj\ or eq.~\hermmode . In table B.1 we list various
operator fields and their hermitean conjugates.
{\topinsert
\setbox\strutbox=\hbox{\vrule height16pt depth3.5pt width0pt} %baselineskip
$$\vbox{\offinterlineskip \halign{\strut#& \vrule# &\hfill~#\hfill
&\hfill~~~~#~\hfill &\vrule#&\hfill#\hfill&\hfill~~~~#~\hfill&\vrule#\cr
\noalign{\hrule}
&\ &\multispan2\hfil $\Phi_{\Delta}$\hfil &\
&\multispan2\hfil $\widehat{\Phi}_{\Delta}$\hfil&\cr
\noalign{\hrule}
&& $X^{\mu}$ &  &&   $X^{\mu}$& &\cr
&& $\del X^{\mu}$& && $-\del X^{\mu}$& &\cr
&& $ e^{i k \cdot X}$& && $e^{-ik\cdot X}$& &\cr
&& $\psi^{\mu}$& && $\psi^{\mu}$& &\cr
&& $\Phi_{(L)}$& $\scriptstyle L=1,\ldots,32$ &&
   $-\Phi_{(L)}$& $\scriptstyle L=1, \ldots, 32$ &\cr
&& $e^{{\Bbb A}_L \Phi_{(L)}}$& $\scriptstyle L=1,\ldots,32$ &&
   $e^{-{\Bbb A}_L \Phi_{(L)}}$& $\scriptstyle L=1, \ldots, 32$  &\cr
&& $\Phi_{(33)}$& && $\Phi_{(33)}$& &\cr
&& $\Phi_{(34)}$& && $\Phi_{(34)} - 2 \log $& &\cr
&& $e^{{\Bbb A}_L \Phi_{(L)}}$& $\scriptstyle L=33,34$ &&
   $e^{{\Bbb A}_L \Phi_{(L)}}$& $\scriptstyle L=33,34$ &\cr
&& $\beta$& && $-\beta$& &\cr
&& $\gamma$& && $\gamma$& &\cr
&& $\eta$& && $-\eta$& &\cr
&& $\xi$& && $-\xi$& & \cr
&& $\del \xi$& && $\del \xi$& &\cr
&& $b$& && $b$& &\cr
&& $c$& && $c$& &\cr
\noalign{\hrule}
}}$$
\centerline{{\bf Table {\bf B1}: } Hermiticity properties of various primary
conformal fields.  }
\sjump
\endinsert}
The cocycle operators $C_{(L)}$ appearing in the expression \spinfield\
and defined in detail in refs.~[\Ref{ammedm},\Ref{mink}] satisfy
$$\eqalignno{ \left( C_{(L)} \right)^{\dagger} & = \
\left( C_{(L)} \right)^{-1} \left( C^{(L)}_{\rm gh} \right)^2 \qquad
{\rm for} \ \ L=1,\ldots,32 & \numali \cr
\left( C_{(L)} \right)^{\dagger} & = \
C_{(L)} \exp \left\{ -2\pi i \sum_{K=1}^{32} Y_{LK} J_0^{(K)} \right\}
\qquad {\rm for} \ \ L=33,34 \ . \cr } $$
Here $C_{\rm gh}^{(L)}$ is the cocycle operator involving the number
operators of the reparametrization ghosts and of the $(\eta,\xi)$ system
(excluding the $(\eta,\xi)$ zero modes)
and is given explicitly by
$$ C_{\rm gh}^{(L)}  = \exp \left\{ -i \pi Y_{34,L} \left(
N_{(\eta,\xi)} - N_{(b,c)} + N_{\bar{b},\bar{c}} \right) \right\} \ .
\efr
Using the hermiticity properties listed above we find that the hermitean
conjugate of the spin field \spinfield\  is given by
$$\eqalignno{ \widehat{S}_{\Bbb A} & = \
\prod_{L=1}^{32} \left( C_{\rm gh}^{(L)} \right)^{2{\Bbb A}_L} \
\exp \left\{ -2\pi i \sum_{L=1}^{32} \left( Y_{34,L} {\Bbb A}_{34}
+ Y_{33,L} {\Bbb A}_{33} \right) J_0^{(L)} \right\} \ \times\cr
&\qquad S_{{\Bbb A}_{34}}^{(34)} S_{{\Bbb A}_{33}}^{(33)}
S_{-{\Bbb A}_{32}}^{(32)} \ldots S_{-{\Bbb A}_{1}}^{(1)} \ .
& \nameali{sahat}\cr }
$$
Here the factors appearing on the right hand side of the equality sign
in the first line of eq.~\sahat\ can be ignored: Acting on any string
state in the theory
they give plus one. It is sufficient to verify this on some generic
ground state $\vert {\Bbb B} \rangle$,
created by the spin field operator
$S_{\Bbb B}$, since non-zero mode creation operators contribute
with integer values to $J_0^{(L)}$. One has to recall that ${\Bbb A}_{33}$
and ${\Bbb A}_{34}$ are always either both integer or both half-integer,
that the number operators appearing in $C^{(L)}_{\rm gh}$ always take
integer values, and finally one has to make use of the fact
(see ref. [\Ref{ammedm}]) that
for a consistent choice of cocycles the following conditions hold
$$\eqalignno{ & \phi_{33} [{\Bbb B}] \equiv \sum_{L=1}^{32} Y_{33,L}
{\Bbb B}_L + \epsilon {\Bbb B}_{33} - Y_{34,33} {\Bbb B}_{34} =\
{\rm integer}\cr
& \phi_{34} [{\Bbb B}] \equiv \sum_{L=1}^{32} Y_{34,L}
{\Bbb B}_L + Y_{34,33} {\Bbb B}_{33} - \epsilon {\Bbb B}_{34} = \
{\rm integer}\cr
& \phi_{33} [{\Bbb B}] \eqmodtwo \phi_{34}[{\Bbb B}]&\numali\cr }
$$
for any ground state $\vert {\Bbb B} \rangle$ existing in the theory,
regardless of the value chosen for the parameter $\epsilon = \pm 1$.

If we finally reorder the individual spin fields in eq.~\sahat\ we obtain
$$\eqalignno{ \widehat{S}_{\Bbb A} & = \ \left( \prod_{L=1}^{32}
\delta_{{\Bbb A}_L + {\Bbb B}_L,0} \right) \left( \prod_{L=33}^{34}
\delta_{{\Bbb A}_L,{\Bbb B}_L} \right)
S^{(1)}_{{\Bbb B}_1} \ldots S^{(34)}_{{\Bbb B}_{34}} \
e^{i \pi {\Bbb B} \cdot Y \cdot {\Bbb B}}\cr
 & = \ \left( \sigma_1^{(33)} {\Bbb C}^{-1} \right)_{{\Bbb A} {\Bbb B}}
S_{\Bbb B} \ , &\numali\cr }
$$
where
$$ \left( {\Bbb C}^{-1} \right)_{{\Bbb A} {\Bbb B}} =
\left( \prod_{L=1}^{33}
\delta_{{\Bbb A}_L + {\Bbb B}_L,0} \right) \delta_{{\Bbb A}_{34},{\Bbb
B}_{34}} \ e^{i \pi {\Bbb B} \cdot Y \cdot {\Bbb B}}
\efr
is the explicit expression for the matrix whose inverse is given by
eq.~\chargeconj .

\appendix{Compatibility of the GSO projection and the map between
${\cal W}_{\vert \lambda \rangle}$ and ${\cal W}_{\langle \lambda \vert}$}
In this appendix we show explicitly that the map \guessthree\
from ${\cal W}_{\vert \lambda \rangle}^{(q)}$ to
${\cal W}_{\langle \lambda \vert}^{(q)}$
is compatible with the GSO projection in the setting of a
four-dimensional KLT heterotic string model [\Ref{KLT},\Ref{Anto}].
In other words we want to show that given a vertex operator
${\cal W}_{\vert \lambda \rangle}^{(q)}$ creating a state
$\vert \lambda \rangle$ satisfying the GSO conditions, the state
$\chi_q \vert \lambda^{\rm BPZ} \rangle$, which is created by
${\cal W}_{\langle \lambda \vert}^{(q)}$,
also satisfies the GSO conditions.

We first recall what is the form of the GSO projection conditions.
We consider
as usual all world-sheet fermions to be bosonized. Then the GSO
conditions involve only the ``momenta'' $J_0^{(L)}$ of the resulting
bosons, and it is sufficient to consider a generic ground state as
created by the operator \groundst . If this state satisfies the GSO
condition, so do all the states obtained from it by means of non-zero
mode creation
operators.

The GSO projection assumes the form (see ref. [\Ref{ammedm}])
$$
{\bf W}_i\cdot {\bf N}_{\modone{{\bfmath\alpha}}} - s_i
(N^{(33)}_{\modone{\alpha_{32}}}
 - N^{(34)}_{\modone{\alpha_{32}}})
- \sum_j k_{ij}m_j - s_i - k_{0i} + {\bf W}_i \cdot
\modone{{\bfmath\alpha}}  \eqmodone 0\ ,
\nfr{GSOp}
where our notation is that of ref.~[\Ref{ammedm}], except for the
labelling of the complex fermions which is chosen in accordance with
ref.~[\Ref{mink}] and the present paper, i.e. the left-moving fermions
are labelled by $L=1, \ldots, 22$, the internal right-moving ones by
$L=23,\ldots,31$, the space-time related ones by $L=32$ (the transverse)
and $L=33$ (the longitudinal), and the superghosts by $L=34$.

Let us briefly recall that the sector (i.e. the set of boundary
conditions for the fermions enumerated by $L=1, \ldots, 32$) is
specified by the 32-component vector
$$ {\bfmath \alpha} = \sum_i m_i {\bfmath W}_i \ , \efr
where the integer $m_i$ takes values $0,1,\ldots,M_i-1$, $M_i$ being the
smallest integer such that $M_j {\bfmath W}_j$ ($j$ not summed) is a
vector of integer numbers. The number operators
$N^{(L)}_{\modone{\alpha_L}}$ are related to the ``momenta'' $J_0^{(L)}
= {\Bbb A}_L$ by
$$\eqalignno{ N^{(L)}_{\modone{\alpha_L}} & = \ {\Bbb A}_L -
\modone{1-\alpha_L} + {1 \over 2} \qquad {\rm for} \quad L=1,\ldots,33
& \nameali{numbop} \cr
N^{(34)}_{\modone{\alpha_{32}}} & = \ {\Bbb A}_{34} -
\modone{\alpha_{32}} - {1 \over 2}  \ . \cr } $$
As we have seen in section 7, given the state $\vert {\Bbb A} \rangle$
with $J_0^{(L)} = {\Bbb A}_L$, created by the spin field operator
\spinfield , the state $\vert {\Bbb A}^{\rm BPZ} \rangle$ created by the
operator \hermground\ has
$$ J_0^{(L)} = \widetilde{{\Bbb A}}_L =
\cases{ -{\Bbb A}_L & for $L=1,\ldots,32$ \cr
+{\Bbb A}_L & for $L=33,34$ \cr}  \ . \nfr{inouttrans}
This behaviour follows directly from the hermiticity properties of the
various fields, as summarized in Appendix B. In eq.~\hermground\ it is
encoded in the presence of the
charge conjugation matrix ${\Bbb C}$ which changes sign on all the
${\Bbb A}_L$ except ${\Bbb A}_{34}$ and the factor $\sigma_1^{(33)}$
which changes sign on ${\Bbb A}_{33}$ only.

Thus we have to check if the GSO projection conditions \GSOp\ are
invariant under the transformation ${\Bbb A}_L \rightarrow
\widetilde{\Bbb A}_L$ given by \inouttrans . The situation is
somewhat complicated by the fact that in general the states $\vert {\Bbb
A} \rangle$ and $\vert {\Bbb A}^{\rm BPZ} \rangle$
do not reside in the same sector.

Let us denote by a
tilde ($\widetilde{\ghost{w}}$) the quantities pertaining to the
state $\vert {\Bbb A}^{\rm BPZ} \rangle$.
We want to show then that if the state $\vert {\Bbb A} \rangle$,
residing in the sector ${\bfmath\alpha}$, satisfies  eq.~\GSOp, then
the state $\vert {\Bbb A}^{\rm BPZ} \rangle$,
residing in the sector $\widetilde{\bfmath\alpha}$,
satisfies
$$
{\bf W}_i\cdot \widetilde{\bf N}_{\modone{\widetilde{\bfmath\alpha}}}
- s_i (\widetilde{N}^{(33)}_{\modone{\widetilde\alpha_{32}}}
 - \widetilde{N}^{(34)}_{\modone{\widetilde\alpha_{32}}})
- \sum_j k_{ij}\widetilde{m}_j - s_i - k_{0i} + {\bf W}_i \cdot
\modone{\widetilde{\bfmath\alpha}}  \eqmodone 0\ .
\nfr{GSOout}
We can actually do something more general, and for this we take the
sum of eqs.~\GSOp\ and \GSOout. We will show that this sum is zero
modulus one, this obviously implies that if eq.~\GSOp\ is satisfied
then also eq.~\GSOout\ is, and viceversa. In summing the two equations
we make use of the fact that the $s_i$ are half-integers~[\Ref{KLT}],
that eq.~\inouttrans\ implies
$N^{(34)}_{\modone{\alpha_{32}}} =
\widetilde{N}^{(34)}_{\modone{\widetilde\alpha_{32}}}$ and
$N^{(33)}_{\modone{\alpha_{32}}} =
\widetilde{N}^{(33)}_{\modone{\widetilde\alpha_{32}}}$ and that the
number operators always have integer eigenvalues. Thus by summing we obtain
$$
{\bf W}_i\cdot \left({\bf N}_{\modone{{\bfmath\alpha}}} +
 \widetilde{\bf N}_{\modone{\widetilde{\bfmath\alpha}}} \right)
- \sum_j k_{ij} \left(m_j + \widetilde{m}_j\right) -2 k_{0i}
+ {\bf W}_i \cdot \left(\modone{{\bfmath\alpha}} +
 \modone{\widetilde{\bfmath\alpha}} \right) \eqmodoneq 0 \ .
\nfr{GSOques}
In order to verify this identity we need to find the relation
between ${\bfmath\alpha}$ and $\widetilde{\bfmath\alpha}$.
We claim that
$$
\widetilde{m}_j\  =\  \cases{ 0         & if $m_j=0$ \cr
                              M_j - m_j & otherwise  \cr} \ ,
\nfr{outsector}
which is consistent with $0\leq m_j,\widetilde{m}_j  \leq M_j - 1$.
The proof is simple: Let $L \in \{ 1, \ldots, 32 \}$. Since the number
operators $N^{(L)}_{\modone{\alpha_L}}$ take integer values, it follows
from eq.~\numbop\ that the allowed values for ${\Bbb A}_L$ in the sector
${\bfmath \alpha} = \sum_j m_j {\bfmath W}_j$ are
$$
{\Bbb A}_{L}\ =\
\frac12 - \sum_j m_j ({\bfmath W}_j)_{(L)} + ({\rm integer}) \ .
\efr
Since
(${{\bfmath W}_j)}_{(L)} = w_{j,(L)}/M_j$ where $w_{j,(L)}$ is an integer
satisfying $0\leq w_{j,(L)}  \leq M_j - 1$, we have
$$
{\Bbb A}_{L} \eqmodone \frac12 - \sum_j m_j {w_{j,(L)}\over M_j}\ .
\nfr{alone}
Likewise, in the sector $\widetilde{\bfmath\alpha}$ we have for
$L=1,\ldots,32$
$$
\widetilde{{\Bbb A}}_{L}\ =\ - {\Bbb A}_{L} \eqmodone \frac12 -
\sum_j \widetilde{m}_j {w_{j,(L)}\over M_j}\ .
\nfr{altwo}
Comparing eqs.~\alone\ and \altwo\ we find the obvious solution
$\widetilde{m}_j = - m_j$  $({\rm mod} M_j)$ which is equivalent to
\outsector .

It is worth noticing that if we sum the two equations
$$
{\Bbb A}_{L}\ \eqmodone\ \frac12 - \modone{\alpha_{L}}
\qquad , \qquad\quad
\widetilde{\Bbb A}_{L}\ =\ - {\Bbb A}_{L}\ \eqmodone\ \frac12 -
\modone{\widetilde{\alpha}_{L}}
\efr
we get
$$
0 \ \eqmodone\ \modone{\alpha_L} +
\modone{\widetilde{\alpha}_L}\ .
\nfr{sumone}
Thus, $\modone{\alpha_{L}} =\modone{\widetilde{\alpha}_{L}}$ only if the
fermion labelled by $L$ satisfies either
Neveu-Schwarz boundary conditions
($\modone{\alpha_{L}} =\modone{\widetilde{\alpha}_{L}} = 1/2$)
or Ramond boundary conditions
($\modone{\alpha_{L}} =\modone{\widetilde{\alpha}_{L}} = 0$).

Let us now return to the identity \GSOques\ that we were supposed to
prove. Substituting eq.~\outsector\ and recalling
that $M_j k_{ij} \eqmodone 0$~[\Ref{KLT}], we find that the term
$\sum_j k_{ij}(m_j + \widetilde{m}_j)$ cancels out. Next we recall
that~[\Ref{KLT}]
$$ 2(k_{0i} + k_{i0}) \eqmodone 2k_{0i} \eqmodone 2 {\bfmath W}_i \cdot
{\bfmath W}_0 \efr
where ${\bf W}_0$ is
the vector with all entries equal to $1/2$.
Substituting all this into eq.~\GSOques\ and using eq.~\numbop\
we find that eq.~\GSOques\ holds if and only if
$$
{\bf W}_i\cdot \left( \modone{{\bfmath\alpha}} +
 \modone{\widetilde{\bfmath\alpha}} -
 \modone{1-{\bfmath\alpha}} - \modone{1-\widetilde{\bfmath\alpha}}
\right) \eqmodoneq 0 \ ,
\nfr{GSOqtwo}
where, rather obviously, $\modone{1-{\bfmath\alpha}}$ is a vector whose
$L$'th component is $\modone{1 - \alpha_L}$. The equation \GSOqtwo\
is indeed satisfied since
$$
\modone{\alpha_L} +
 \modone{\widetilde{\alpha}_L} -
 \modone{1-\alpha_L} -
\modone{1-\widetilde{\alpha}_L} \ =\ 0\ .
\nfr{GSOqthree}
Indeed, if $\modone{\alpha_L}=0$ then also
$\modone{\widetilde{\alpha}_L} =
 \modone{1-\alpha_L} =
\modone{1-\widetilde{\alpha}_L} = 0$ and eq.~\GSOqthree\ is trivially
satisfied. Otherwise
$\modone{1-\alpha_L} = 1 - \modone{\alpha_L}$,
$\modone{1-\widetilde{\alpha}_L} = 1 -
\modone{\widetilde{\alpha}_L} $ and by eq.~\sumone\
$\modone{\alpha_L} + \modone{\widetilde{\alpha}_L} = 1$ and again
eq.~\GSOqthree\ holds.

Thus eq.~\GSOques\ is satisfied, and we have shown that if the
vertex operator ${\cal W}_{\vert \lambda \rangle}^{(q)}$
creates a state in the GSO projected spectrum,
i.e. a state that satisfies eq.~\GSOp, then so does the
vertex operator ${\cal W}_{\langle \lambda \vert}^{(q)}$, and viceversa.
\references
\beginref
\Rref{KLT}{H.~Kawai, D.C.~Lewellen and S.-H.H.~Tye, Nucl.Phys.
{\bf B288} (1987) 1.}
\Rref{Verlinde}{E.~Verlinde and H.~Verlinde, Phys.Lett. {\bf B192}
(1987) 95.}
\Rref{PDV1}{P.~Di Vecchia, M.L.~Frau, K.~Hornfeck, A.~Lerda, F.~Pezzella
and S.~Sciuto, Nucl.Phys. {\bf B322} (1989) 317.}
\Rref{PDV2}{P.~Di Vecchia, invited talk at the Workshop on
String quantum Gravity and Physics at the Planck scale, Erice,
June 1992, Int.Journ.Mod.Phys. {\bf A}.}
\Rref{Anto}{I.~Antoniadis, C.~Bachas, C.~Kounnas and P.~Windey,
Phys.Lett. {\bf 171B} (1986) 51; \newline
I.~Antoniadis, C.~Bachas and C.~Kounnas, Nucl.Phys. {\bf B289} (1987) 87;
\newline I.~Antoniadis and C.~Bachas, Nucl.Phys. {\bf B298} (1988) 586.}
\Rref{FMS}{D.~Friedan, E.~Martinec and S.~Shenker, Nucl.Phys. {\bf B271}
(1986) 93.}
\Rref{Sonoda}{H.~Sonoda, Nucl.Phys. {\bf B326} (1989) 135.}
\Rref{Berera}{A.~Berera, Nucl.Phys. {\bf B411} (1994) 157.}
\Rref{Weinberg}{S.~Weinberg, Phys.Lett. {\bf B156} (1985) 309.}
\Rref{Hoker}{K.~Aoki, E.~D'Hoker and D.H.~Phong, Nucl.Phys. {\bf B342}
(1990) 149;\newline
E.~D'Hoker and D.H.~Phong, Phys.Rev.Lett. {\bf 70} (1993) 3692,
Theor.Math.Phys. {\bf 98} (1994) 306 \hbox{hep-th/9404128},
Nucl.Phys. {\bf B440} (1995) 24 \hbox{hep-th/9410152}.}
\Rref{Weisberger}{J.L.~Montag and W.I.~Weisberger, Nucl.Phys. {\bf B363}
(1991) 527.}
\Rref{GSW}{M.B.~Green, J.H.~Schwarz and E.~Witten, {\sl Superstring Theory},
Cambridge University Press, 1987.}
\Rref{Phong}{For a review, see
E.~D'Hoker and D.H.~Phong, Rev.Mod.Phys. {\bf 60} (1988) 917.}
\Rref{Kaj}{G.~Cristofano, R.~Marotta and K.~Roland, Nucl.Phys. {\bf
B392} (1993) 345.}
\Rref{Bluhm}{R.~Bluhm, L.~Dolan and P.~Goddard, Nucl.Phys. {\bf B309}
(1988) 330.}
\Rref{Kaj2}{K.~Roland, Phys.Lett. {\bf B312} (1993) 441.}
\Rref{Veneziano}{D.~Amati, M.~Ciafaloni and G.~Veneziano, Phys.Lett.
{\bf B197} (1987) 81.}
\Rref{IZ}{C.~Itzykson and J.-B.~Zuber, ``{\sl Quantum Field Theory}",
McGraw-Hill, New York, 1980.}
\Rref{ammedm}{A.~Pasquinucci and K.~Roland, ``{\sl On the computation
of one-loop amplitudes with external fermions in 4d heterotic
superstrings\/}'', Nucl.Phys. {\bf B440} (1995) 441,
\hbox{hep-th/9411015}.}
\Rref{mink}{A.~Pasquinucci and K.~Roland, ``{\sl Bosonization of world-sheet
fermions in Minkowski space-time\/}", Phys.Lett. {\bf B351} (1995) 131,
\hbox{hep-th/9503040}.}
\Rref{Zwiebach}{B.~Zwiebach, Nucl.Phys. {\bf B390} (1993) 33,
\hbox{hep-th/9206084}.}
\Rref{paolosewing}{P.~Di Vecchia, in DST Workshop on Particle Physics
--Superstring Theory, eds. H.S.~Mani and R.~Ramachandran, World
Scientific, p.41.}
\Rref{BK}{Z.~Bern and D.~Kosower, Nucl.Phys. {\bf B379} (1992) 451.}
\Rref{Marco}{G.~Cristofano, M.~Fabbrichesi and K.~Roland, Phys.Lett.
{\bf B244} (1990) 397; \newline
A.~Bellini, G.~Cristofano, M.~Fabbrichesi and K.~Roland, Nucl.Phys. {\bf
B356} (1991) 69.}
\Rref{Paolo}{P.~Di Vecchia, M.~Frau, K.~Hornfeck, A.~Lerda, F.~Pezzella
and S.~Sciuto, Nucl.Phys. {\bf B333} (1990) 635.}
\Rref{BPZ}{A.A.~Belavin, A.M.~Polyakov and A.B.~Zamolodchikov, Nucl.Phys.
{\bf B241} (1984) 333.}
\endref
\ciao
